\documentclass[12pt]{article}

\usepackage{graphicx}
\usepackage{float}
\usepackage{setspace}
\usepackage{amsfonts}
\usepackage{amsmath}
\usepackage{amssymb}
\usepackage{hyperref}
\usepackage{url}
\usepackage{xurl}
\usepackage[dvipsnames]{xcolor}
\usepackage{tikz}
\usepackage{epigraph} 
\usepackage{booktabs}

\usepackage{soul}
\usepackage{threeparttable}
\usepackage{caption}
\usepackage{multirow}
\usepackage{subfig}
\usepackage[utf8]{inputenc}
\usepackage{placeins}
\usepackage{sverb}
\usepackage{subcaption}
\usepackage{etoc} % for tables of contents
\usepackage{makecell} % for table column headers
\usepackage{longtable}
\usepackage{threeparttablex}
\usepackage{booktabs}
\usepackage{siunitx}

\hypersetup{
    colorlinks=true,
    linkcolor=black,
    filecolor=cyan,      
    urlcolor=black,
    citecolor=.}
    
\usepackage[symbol]{footmisc}

\usepackage{geometry}
\usepackage{natbib}
\usepackage{har2nat}
\usepackage{authblk}

\DeclareMathOperator{\logit}{logit}

\usepackage{multibib}
\newcites{App}{Appendix References}

\makeatletter
\renewcommand{\l@section}{\@dottedtocline{1}{1.5em}{2.6em}}
\renewcommand{\l@subsection}{\@dottedtocline{2}{4.0em}{3.6em}}
\makeatother
\title{\bf{Are Politicians Responsive to Mass Shootings? Evidence from U.S. State Legislatures}\thanks{\scriptsize{The authors contributed equally and are listed alphabetically. We thank Emily Rusting and Luis Muñoz for their excellent research assistance, the Giffords Law Center for providing access to internal data, as well as the Violence Project and Gun Violence Archive for data on mass shootings. We are also indebted to the participants of the qualitative interviews reported in this paper, and thank them for their time and public service. For helpful discussion and comments, we thank Eric Baldwin, Kathy Bawn, Ryan Baxter-King, Nathaniel Birkhead, John J. Donohue, Chad Hazlett, Takuma Iwasaki, Justin Kirkland, Matthew Lacombe, Adam Lankford, Jeffery Lewis, Amanda Mauri, Julia Payson, Chris Tausanovitch, Daniel Thompson, Lynn Vavreck, attendees of the 2024 \& 2025 American Political Science Association Annual Meeting, the 2025 Conference for Empirical Legal Studies, and the 2024 National Research Conference for the Prevention of Firearm-Related Harms, members of the Practical Causal Inference Lab, and members of the UCLA American Politics Working Group. Special thanks to the UCLA Initiative to Study Hate for their generous support of this research.}}}

\author{
Haotian Chen\footnote{\scriptsize{PhD Candidate, Department of Political Science, UCLA. barneychen@ucla.edu, \url{https://www.haotianchen.com}.}} \\ \vspace{-3mm}
Jack Kappelman\footnote{\scriptsize{PhD Candidate, Department of Political Science, UCLA. jack.kappelman@gmail.com, \url{https://www.jakappelman.com}.}}
\
}

\IfFileExists{upquote.sty}{\usepackage{upquote}}{}

\begin{document}

% Set up a non-existent table of contents so it starts at the right point later
\etocdepthtag.toc{mtchapter}
\etocsettagdepth{mtchapter}{section}
\etocsettagdepth{mtappendix}{none}

\maketitle

\thispagestyle{empty}
\linespread{1}

\vspace{5mm}

\begin{abstract}

The United States leads the world in the number of mass shootings that occur each year, even as policy making on firearms remains polarized along party lines.
In the face of increasing violence and public demand for policy action, we ask whether legislators change their voting behavior on firearm policy in the wake of mass shootings.
We estimate the latent gun-policy positions of 14,585 state legislators across all 50 states using roll-call votes on firearm-related bills from 2011 to 2022. Employing a difference-in-differences design, we find that mass shootings occurring within a legislator’s district do not, on average, measurably shift their positionality on firearm policy. This null effect is robust across analyses accounting for legislators’ partisanship, their geographic proximity to the shooting, and characteristics of individual shootings. 
Our findings suggest that even acute, locally salient tragedies fail to cause changes in how legislators vote on firearm policy.

\end{abstract}

\vspace{20mm}

\pagebreak
\setcounter{page}{1}

\begingroup
\setstretch{1.5}

\section{Introduction}
On December 14th, 2012, a mass shooter entered Sandy Hook Elementary School in Newtown, Connecticut, and fatally shot six staff and twenty children between six and seven years old. The shooting, like many that came before and many that have since occurred, sparked national discussions about firearm policy, with President Barack Obama declaring in a press conference on the day of the shooting that American lawmakers needed to ``come together and take meaningful action to prevent more tragedies like this, regardless of the politics.''\footnote{See \href{https://obamawhitehouse.archives.gov/photos-and-video/video/2012/12/14/president-obama-makes-statement-shooting-newtown-connecticut}{``Statement by the President on the School Shooting in Newtown, CT,'' \textsc{Obama White House Archives}.}} Back in Connecticut, two Republican politicians represented the state legislative districts surrounding Sandy Hook. Mitch Bolinsky, a freshman lawmaker in the Connecticut General Assembly, was profiled by local media after the shooting for breaking ranks with the state's Republican delegation and voting in favor of what was then regarded as ``the nation's toughest gun-control law.''\footnote{See \href{https://www.newstimes.com/local/article/a-freshman-lawmaker-s-year-of-crisis-4602792.php}{``A freshman lawmaker's year of crisis,'' \textsc{The News-Times}.}} John McKinney, the state senator representing Newtown and the Minority Leader of the Connecticut Senate, reportedly told his caucus after the shooting that he intended to ``go negotiate and work with the Democrats,''\footnote{See \href{https://www.cnn.com/2022/12/14/us/sandy-hook-10-year-anniversary/}{``Sandy Hook parents push for changes in the decade since the school shooting,'' \textsc{CNN}.}} and later played a substantial role in enacting the same law that Representative Bolinsky broke with his party to vote for. Yet, these responses may be exceptional, and beg the question: do mass shootings cause legislators to alter their voting behaviors on firearm policy?

This puzzle is especially acute because gun policy enjoys bipartisan support among the public. For decades, pluralities of Americans have supported restrictive policies \citep{schuman1977attitude}, with Gallup polls suggesting that in 2022, as many as 92\% of U.S. adults, 97\% of Democrats, and 90\% of Republicans, favored background checks for all gun sales.\footnote{See \href{https://news.gallup.com/poll/394022/public-pressure-gun-legislation-shootings.aspx}{``Public Pressure for Gun Legislation Up After Shootings,'' \textsc{Gallup}}.} Between 2012 and 2022, the period we study in this paper, Americans' dissatisfaction with U.S. gun laws rose from 42\% to 56\%.\footnote{See \href{https://news.gallup.com/poll/470588/dissatisfaction-gun-laws-hits-new-high.aspx}{``Dissatisfaction With U.S. Gun Laws Hits New High,'' \textsc{Gallup}.}} And yet, policymaking remains deeply polarized, and federal action on firearm policy has been effectively frozen for decades \citep{spitzer_politics_gun_2023}. Responses at the state level (where legislatures debate hundreds of gun bills annually) even appear decoupled from public preferences, and legislating after mass shootings appears to be driven by partisan control rather than majoritarian preferences; the directionality of a states' gun policy environment following such shootings appears to follow partisan patterns \citep{luca_impact_2020}. Why does it seem that high degrees of public support, coupled with the salience of a focal tragedy, fail to produce legislative action that addresses the violence experienced by the very communities that elected officials represent?

Extant scholarship offers competing predictions. Accountability theories suggest that legislators ought to respond to local crises because voters appear to hold governmental institutions and elected officials at least somewhat responsible when disasters occur under their watch \citep[e.g.,][]{fiorina_economic_1978, healy_myopic_2009, gasper_make_2011, heersink_disasters_2017}. This logic explains legislators pursuing tougher crime bills after spikes in violent crime \citep{arnold_holding_2012}, protectionist trade votes following economic shocks \citep{feigenbaum2015legislators}, and more support for emergency aid after natural disasters \citep[e.g.,][]{arceneaux_who_2006, bechtel_how_2011, malhotra_attributing_2008, atkeson_catastrophic_2012, carlin_natural_2014}. From this perspective, mass shootings, which are highly salient, emotionally charged, and geographically concentrated, should create strong incentives for responsive actions among legislators.

However, theories of partisan polarization suggest that these mechanisms of accountability may fail in highly polarized domains \citep{NBERw20253, butler_limits_2018}. To the degree that mass shootings or exposure to gun violence mobilizes voters \citep[e.g.,][]{shoub2025effect, markarian2024exposure, hassell_mobilize_2020} or shifts localized public opinion \citep{gunn_online_2018, barney_reexamining_2019, newman_mass_2019, rogowski_critical_2019, sharkey_effect_2021, Baxter-King2024-fr}, legislators may lack incentive to respond if deviating from party orthodoxy invites primary challenges \citep{NBERw20253}, or unwanted attention from interest groups \citep{baldwin2025gunPACs} and vocal constituencies \citep{lacombe_gun_2019, lacombe_political_2019}. These mechanisms may insulate voting behavior on gun policy from district-level shocks, even when legislators are personally affected.

This tension --- between accountability pressure and partisan constraint --- is our focus. It is possible that legislators may be more responsive to gun violence as gun violence intensifies. Annual firearm fatalities in the U.S. increased by 45.5\% between 2004 and 2021 \citep{rees2022trends}. Mass shootings have become both more frequent and more lethal, with several of the deadliest shootings in U.S. history occurring during our study period \citep{peterson2024epidemiology}. To the degree that public pressure could overpower partisan gridlock on gun policy, this escalation may amplify that pressure. Mass shootings generate intense media coverage \citep{schildkraut2018mass}, mobilize advocacy groups \citep{baldwin2025gunPACs}, and increase public engagement with gun policy \citep{reny_public_2023}. When such violence occurs in a legislator's own district, it should create the maximum pressure needed for a detectable response. The question is whether this pressure, even when acute and local, is sufficient to dislodge entrenched partisan positions.

We address this question by examining state legislative voting behavior across the United States. We combine roll-call records with a novel dataset from the Giffords Law Center (GLC) that ranks over 15,000 firearm-related bills introduced between 2011 and 2022 on a permissive-to-restrictive scale. Using pairwise comparison, we construct an annual Gun Issue Score for 14,585 state legislators across the United States over a 12 year period. This metric captures firearm-policy-specific ideal points that are comparable across states and over time.

Through a difference-in-differences design accounting for heterogeneity in treatment timing, we estimate whether the occurrence of a mass shooting within (or near) a legislator's district causes a measurable change in their position on gun policy. We focus our analysis on estimating the aggregate effect of mass shootings rather than shooting-specific effects. This approach boosts statistical power by pooling rare events, yields an informative estimate of typical responses, avoids multiple comparisons, and uses methods accounting for heterogeneity to identify a single average treatment effect. Across 53 mass shootings affecting 90 legislators in 23 states, we find that such events have an estimated effect near zero; relative to cross-legislator variation in Gun Issue Scores, the 95\% confidence interval excludes effects larger than roughly ±0.2 standard deviations. 

This finding conflicts with the self-reported assessments of responsiveness that California state legislators described in qualitative interviews we conducted, which would suggest that our findings ought to instead illustrate shifts in legislative behavior. However, these legislators also suggested that even when personally confronted with a mass shooting in their district, they perceived substantial constraints on their ability to deviate from party-aligned positions, reinforcing a theory-driven expectation of null effects. In sum, our mixed-methods approach finds that legislators do not change their voting behavior following a mass shooting, though this finding only speaks to a single metric of legislator behavior. Substantively, we establish boundaries of democratic responsiveness: if legislators do not respond to acute local tragedies even as violence escalates, what shocks --- if any --- can disrupt legislative inertia?

\section{Should Mass Shootings Cause Changes in Legislator's Behavior?}
Our focus on state legislators is motivated by the contemporary landscape of firearms policymaking in the United States. With federal gun policy largely stagnant \citep{spitzer_politics_gun_2023}, state legislatures have become the primary venues for firearm debates \citep{luca_impact_2020}. Measuring issue-specific positionality via roll-call behavior requires a high-density voting environment: while Congress rarely votes on gun-related bills, state legislatures produce tens of thousands of such records. If mass shootings cause legislators to change their voting behavior on gun policy, it should be most observable at the state level, where the bulk of modern firearms policymaking occurs.

State legislatures thus provide an ideal setting to test theories of legislative responsiveness. To clarify why mass shootings may fail to shift voting behavior, we draw on qualitative interviews with five California state legislators who represented districts where mass shootings occurred (2010–2022), detailed in Appendix \ref{app:interview}. Under longstanding theories of legislative responsiveness to contemporaneous events, one might expect that mass shootings serves as focusing events \citep{birkland_after_1997} or windows of policy opportunity \citep{kingdon_agendas, baumgartner_agendas_2010} that allow legislators to prioritize specific interventions. Given that annual firearm deaths in the United States increased, on average, across our period of study \citep{rees2022trends}, the increasing magnitude and severity of gun violence may put more pressure on political elites to act. Legislator A (D) reflected on this, noting that a shooting in their district provided an opportunity to ``use the moment... to also give [the victims] lives and their deaths meaning,'' while Legislator B (D) emphasized that if advocates ``don't try to act in the aftermath of a horrific tragedy, [they] have... lost an opportunity for public attention to be focused for long enough to get something to actually change.'' If exposure to gun violence mobilizes the electorate --- and recent work suggests that it does \citep{shoub2025effect, markarian2024exposure} --- legislators may be incentivized to change their voting behavior on gun policy to match the preferences of their constituencies. 

However, it is also possible that partisan polarization on gun policy places constraints on the behaviors of legislators that mass shootings are not sufficiently capable of dislodging. In the cycle of ``outrage-action-reaction'' that characterizes the politics of gun policy in the United States \citep{spitzer_politics_gun_2023}, Democrats face constant pressure to restrict access to firearms while Republicans face pressure to defend gun rights. This cyclical nature ensures a stagnant policy environment, especially given the polarized nature of state legislatures during this period \citep{rogers2017electoral, phillips_who_runs_for_congress_2022, shor2011ideological}. For legislators already at a party's ideological extreme on gun policy, further movement in their party's preferred direction may be constrained by ceiling effects, while movement toward the opposing party's position would require overcoming substantial partisan barriers. However, not all legislators occupy these extreme positions, so the strength of such constraints may vary across contexts, which could result in heterogeneous treatment effects.

Additionally, party discipline or interest-group pressure, for example, may discourage deviation even among legislators with some ideological room to move. Gun rights advocates maintain exceptionally high levels of political organization \citep{lacombe_political_2019}, pay close attention to policy shifts \citep{lacombe_gun_2019}, and legislators on either side of the issue benefit from significant industry-related campaign contributions \citep{baldwin2025gunPACs, richards_guns_2017}. Legislators risk alienating these powerful constituencies and jeopardizing their reelection chances if they break with party orthodoxy \citep{NBERw20253}. Legislator C (R) described a direct threat from party leadership following a shooting in their district: ``I was told directly, if I speak about [gun control], it's going to cost \$1,000,000 per Tweet. For us to defend you in the next election, it's going to cost 1,000,000 bucks a Tweet.''\footnote{We cannot confirm the veracity of Legislator C's claims about sanctions threatened by the California GOP. We can, however, confirm that Legislator C did not post a single Tweet related to a firearms policy proposal until 181 days after the shooting. This Tweet was expressing opposition to a permissive policy.} Such constraints may prevent legislators from translating personal reactions into changes in voting behavior, even when they are not at the ideological limit.

These pressures help explain why legislative behavior on firearms policy remains remarkably inert following shootings; together, such constraints ensure that voting behavior is largely insulated from district-level shocks, even when legislators are personally affected by incidents in their district. While legislators may respond symbolically --- through vigils, meetings, or public statements --- we expect to find that voting behavior is anchored by partisan fidelity. In this paper, we ask whether incidents of mass firearm violence, which may serve as focal points for constituencies and --- under theories of democratic responsiveness --- their representatives, cause legislators for whom these events are most salient to act differently from their peers. 

\section{Data and Methods}

\subsection{Mass Shootings in Scope of Study}

To identify mass shootings for the purposes of this study, we rely upon a definition established by the Congressional Research Service, which is aligned with definitions used by researchers across a variety of disciplines \citep[e.g.,][]{luca_impact_2020, violence_project}. Under this definition, a mass shooting is an incident in which four or more victims (not including the shooter) are murdered with firearms in a single, continuous incident in a public location that is unrelated to other types of criminal activity (such as gang violence).\footnote{See \href{https://sgp.fas.org/crs/misc/R44126.pdf}{``Mass Murder with Firearms: Incidents and Victims, 1999-2013,'' \textsc{Congressional Research Service}}. Non-mass shootings are nearly universal across districts, limiting identification (95 percent of lower and 99 percent of upper chamber districts experienced at least one shooting between 2019–2025; Figure \ref{fig:gva_map}). Mass shootings are rarer and spatially concentrated, preserving a credible treated–untreated comparison.} 

We consider all mass shootings that occur in the United States between 2011 and 2022 as within our scope of study, yielding a total of 68 mass shootings in 29 states. However, due to limitations in voting data and our identification requiring at least two pre-treatment and one post-treatment observation for at least one affected legislator, our sample of shootings is reduced to encompassing 53 incidents in 22 states. Additional details on these shootings and the selection decisions underlying the inclusion of these events in our analyses can be found in Appendix \ref{app:shootings}.

\subsection{Measuring Legislators' Gun Policy Ideal Points}

How polarized have legislators become on gun policy, and to what degree do mass shootings lead them to adjust their voting behavior on firearm-related bills? To answer this question, we first turn to a dataset maintained by the Giffords Law Center (GLC), a gun violence prevention advocacy organization.

\subsubsection{GLC Firearm Bill Rankings and Bill Effects}

From 2011 to 2023, the GLC hand-coded every bill introduced or enacted by every state legislature that was directly related to firearms policy, with the universe of legislation ranging from provisions allowing the permitless carry of concealed firearms, to large-scale budgetary items mentioning firearms in a single sub-item, to bans on assault-style weapons. This yields a total of 15,299 legislative attempts at the state level. For each of them, the GLC ranked it on a ``weaken/neutral/strengthen'' scale with regards to the bill's likely effect on a state's respective gun control policy environment (with ``strengthen'' bills making a state policy more restrictive and therefore more favorable to gun violence prevention advocates, for instance).\footnote{Under the terms of a non-disclosure agreement dictating the amount and type of data that we are able to share in replication materials and publications, we arranged for conditional access to this as-of-yet untapped source of internal lobbying data.} 

In section \ref{app:GLC} of the Appendix, we discuss our validation procedure used for the GLC Bill Rankings, and we ultimately accept that ``weaken'' laws (according to the GLC) can generally be considered permissive (-1), ``neutral'' as neutral (0), and ``strengthen'' laws as restrictive (+1). We refer to this -1/0/+1 score as the ``Bill Effect'' throughout this paper --- a novel measure of the relative restrictiveness or permissiveness of a firearm-related legislative measure introduced in a state. Table \ref{tab:bill_examples} shows three examples of bills from California during the period that we study, with the original GLC ranking and the Bill Effect.

\subsubsection{Measuring Legislators' Latent Position on Firearm-related Issues}
\label{score_method}

Political Scientists have developed various methods of measuring a legislator's ideal points through their roll-call votes \citep[e.g.,][]{poole1997congress, ansolabehere_roll_call_2001, shor2011ideological}. To measure an individual legislator's position on firearm-related policies over time, we develop an issue-specific ideal point ($\lambda$), which we refer to as the ``Gun Issue Score,'' by combining the Bill Effects and the roll-call vote data for both chambers of the state legislatures:\footnote{The California Assembly and Senate roll call vote data is accessed through California Legislative Information website (\url{https://leginfo.legislature.ca.gov}). Roll call voting data for other states is collected from LegiScan (\url{https://legiscan.com}).} 

For state legislator $i$, let $\lambda_i$ be their latent gun control ideal points in a given year. We first map each vote on firearm-related legislation in the direction of supporting restrictive policies (support for gun control) for direct comparison (including floor votes among the entire chamber and committee or procedural votes where available). We then employ the Bradley–Terry model \citep{bradley1952rank, hopkins2022trump} so that in each of the pairwise comparisons between legislator $i$ and legislator $j$, the probability of $i$ being more supportive of restrictive policies than $j$ is $p_{ij}$. Then, the log-odds corresponding to $p_{ij}$ is:
\begin{equation}
    \logit(p_{ij}) =  \lambda_i - \lambda_j
\end{equation}

Assuming that each pairwise comparison is independent, the latent trait $\lambda$ can be estimated by maximum likelihood. As we wish to compare legislators across time, we anchor each state-year's raw gun issue scores to a stable reference metric: the \citet{shor2011ideological} state-legislator ideology score. In each state-year, we use legislators who are observed in both datasets as bridging actors to pin down the zero point and units of our gun score, estimating a robust linear link from our raw gun scale onto the Shor–McCarty scale and then applying that same link to all legislators in that state-year. The assumption is that the Shor–McCarty scale provides a consistent reference metric over time within a state, so this anchoring makes our gun scores comparable across years within a state.\footnote{See \citet{bailey2007comparable, lewis2015does, lewis2020estimating} for discussions of limitations on the comparability of ideal point estimates.} Unlike the conventional methods of using a spatial choice model or Bayesian Item Response Theory (IRT) model on a subset of roll-call voting records, our scoring regime uses the direct Bill Effects to orient scores in the direction of supporting gun control policies. As robustness checks, we also estimate latent traits using the IRT model \citep{clinton2004statistical} in \ref{app:IRTresults} and discuss alternative measures such as survey responses in \ref{app:NRAscores}. 

Figure \ref{fig:gun_policy_time} shows the distribution of the Gun Issue Scores over time for U.S. state legislators. We pick up a clear separation between Democratic and Republican legislators (left panels), suggesting that partisan legislators are polarized on this policy domain. We pick up a small increase in the separation between the parties over time (top left panel). 
The bottom left panel shows the density distribution of all legislators, by party, across all sessions. The right hand panel calculates the difference between Republican and Democratic median Gun Issue Scores within each state (over the entire study period), and ranks states by this metric, which captures state-level polarization relative to the difference in party medians nationally. Clearly, California is the most polarized state for the gun policy domain. We find that the most permissive legislator during our study period is John Tobia (R-FL);\footnote{Tobia is was notably the only member among 160 Florida Representatives and Senators to vote against a gun control measure introduced after the Sandy Hook school shooting. See \href{https://www.gainesville.com/story/news/state/2013/06/21/nra-catches-heat-for-backing/31848986007/}{``NRA catches heat for backing Fla. gun law,'' \textsc{The Gainseville Sun}}.} the most restrictive legislator is Daniel Hernandez (D-AZ).\footnote{Hernandez was formerly an intern for U.S. Representative Gabby Giffords, and was present at the 2011 constituent meeting when she was shot. He has been credited with saving her life. See \href{https://www.cnn.com/2013/02/05/living/daniel-hernandez-giffords-book/index.html}{``Intern’s memoir recalls Giffords shooting,'' \textsc{CNN}}.}

\begin{figure}[!h]
    \centering
    \caption{\textbf{Polarization on Firearms Policy in U.S. State Legislatures, 2011-2022}}
    \includegraphics[scale=.85]{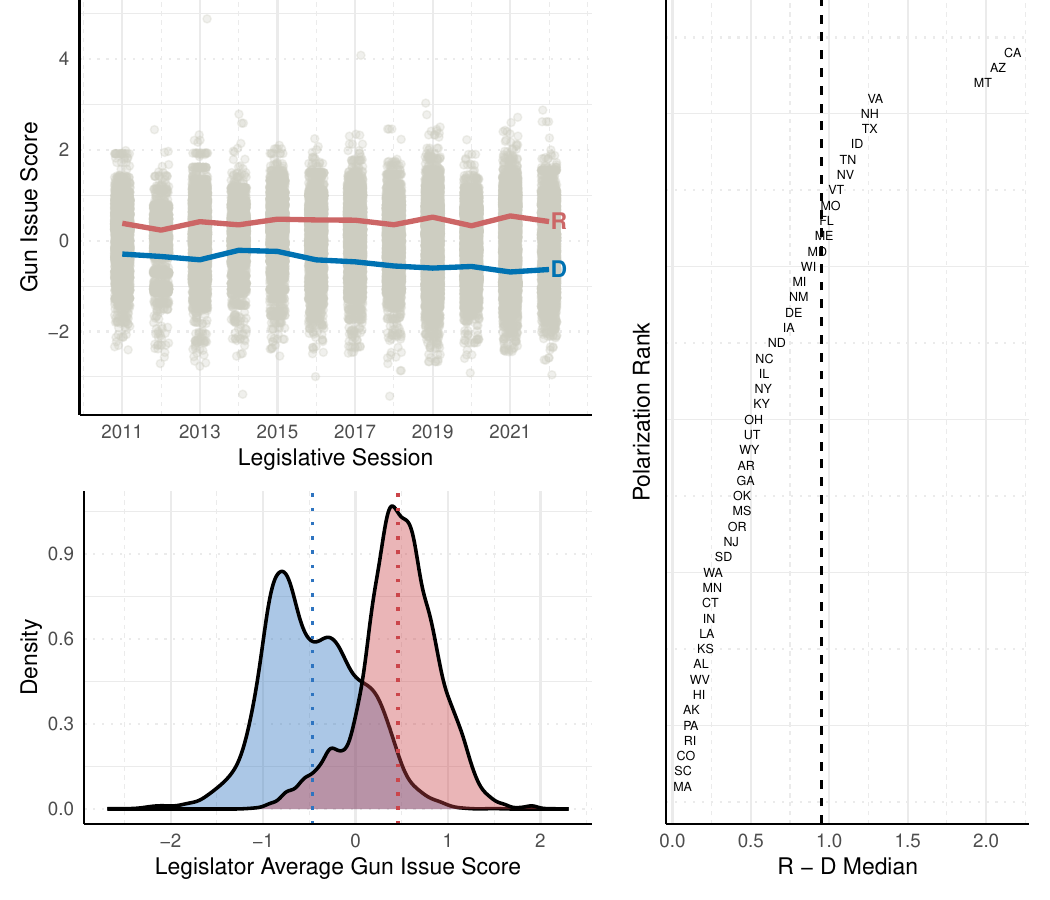}
    \begin{flushleft}
    \scriptsize{\textit{Note}: The top-left panel plots legislator-specific ideal points over time, and shows cross-state party averages across years as solid, labeled lines. The bottom-left panel displays density estimates of legislator-level average Gun Issue Score scores by party ({\color{MidnightBlue}blue} = Democrats, {\color{BrickRed}red} = Republicans) across all states and sessions, with vertical dotted lines indicating the averages of each party’s distribution. The right panel ranks states by polarization, defined as the difference between Republican and Democratic median Gun Issue Score scores within each state (over the entire study period). The vertical dashed line denotes the national polarization level, calculated as the difference between party medians pooled across all states. States to the right of this line exhibit greater polarization than the national median, while states to the left are less polarized. Data for Nebraska is dropped from the bottom-left and right panels due to all state legislators being registered as non-partisans.}    
    \end{flushleft}
    \label{fig:gun_policy_time}
\end{figure}

\subsection{Treatment Assignment and Identification}

As seen in Figure \ref{fig:affected_districts_US} and Table \ref{tab:shootings_in_study}, the shootings we consider as treatments occurred across different legislative districts, states, and time periods. Given this heterogeneity in treatment timing, traditional two-way fixed effects estimators are inappropriate, as they fail to account for variation in treatment effects due to staggered adoption \citep{goodman_did_treatment_timing_2021}. We therefore employ a stacked difference-in-differences (DiD) estimation strategy following \citet{cengiz_stacked_2019}.

Our approach rests on two key assumptions. First, treatment is absorbing: once a legislator experiences a mass shooting in or near their district, they remain treated throughout the observation period. Second, we assume parallel trends with never-treated units: legislators who never experience a shooting in their district during our study period serve as controls and would have followed similar voting trajectories absent treatment. We validate these assumptions through pre-trend analysis in Section \ref{app:parallel_trends_pre_trends}.

For treatment assignment, shootings occurring in the first six months of a year render that year post-treatment, while shootings in the last six months render the following year post-treatment. In states where the legislature meets biennially (such as Texas), the session immediately following a shooting is designated as the first post-treatment period. We require at least two pre- and one post-shooting observation per legislator for identification, so we drop those whose tenure begins in the first post-treatment period or ends in the final pre-treatment period. See Section \ref{app:treat_timing} for additional details --- in total, 41 legislators are dropped because of failure to meet these criteria, leaving 90 legislators for analysis in our main specification.

We construct shooting-specific datasets, one per shooting $g$, that include treated legislators and all never-treated units, excluding legislators treated at other times.\footnote{Each shooting ($g$) occurs in a single state, so the index $g$ implicitly captures both the shooting and its state. Our modeling structure does not account for spillover effects across state lines.} These datasets are stacked, and we estimate the following two-way fixed effects specification:
    \begin{align} 
    y_{it} = \alpha_{ig} + \omega_{ig} + \gamma_{tg} + \overset{-2}{\underset{k=-K}{\sum}}\mu_k D_{it}^k + \overset{L}{\underset{k=0}{\sum}}\mu_k D_{it}^k + \epsilon_{it} 
    \end{align}
where \(y_{it}\) denotes the Gun Issue Score for legislator \(i\) in year \(t\). We utilize relative-time indicators (\(D_{it}^k = \mathbb{I}[t-g_i = k]\)) rather than a binary treatment indicator, allowing us to estimate dynamic treatment effects across multiple pre- and post-treatment periods. The first summation captures lead periods, while the second captures treatment effects, with $k=-1$ (the period immediately before treatment) excluded as the reference category. The coefficients $\mu_k$ represent the average treatment effect on the treated (ATT) at $k$ periods from the shooting, relative to the excluded period. 

We include  legislator- (\(\alpha_{ig}\)), chamber- (\(\omega_{ig}\)), and year-fixed effects (\(\gamma_{tg}\)). Legislator-fixed effects absorb all time-invariant characteristics of individual legislators, including any time-invariant state-level attributes, such that separate state-fixed effects are not separately identified in this notation (even though they are included in our models). Year-fixed effects account for common shocks across all legislators within a given period, while chamber-fixed effects flexibly control for persistent differences between legislative chambers within states. Robust standard errors are clustered at the state, chamber, legislator, and year levels to account for serial correlation and cross-sectional dependence across these dimensions.\footnote{Multi-way clustering accounts for potential error correlation within legislators over time, within chambers and states due to shared institutional environments, and across legislators within years due to common shocks. Failure to account for these dependencies can bias standard errors downward.}

\section{Results}
\subsection{Shootings Occurring Within or Near a State Legislator's District Do Not Affect Their Position on Firearm-Related Policies}

\begin{figure}[h!]
    \centering
    \caption{\textbf{ATT Estimates for Effect of Mass Shooting on Gun Issue Score Demonstrate Null Results, Across Party, for All Shootings in Study}}
    \includegraphics[scale=1]{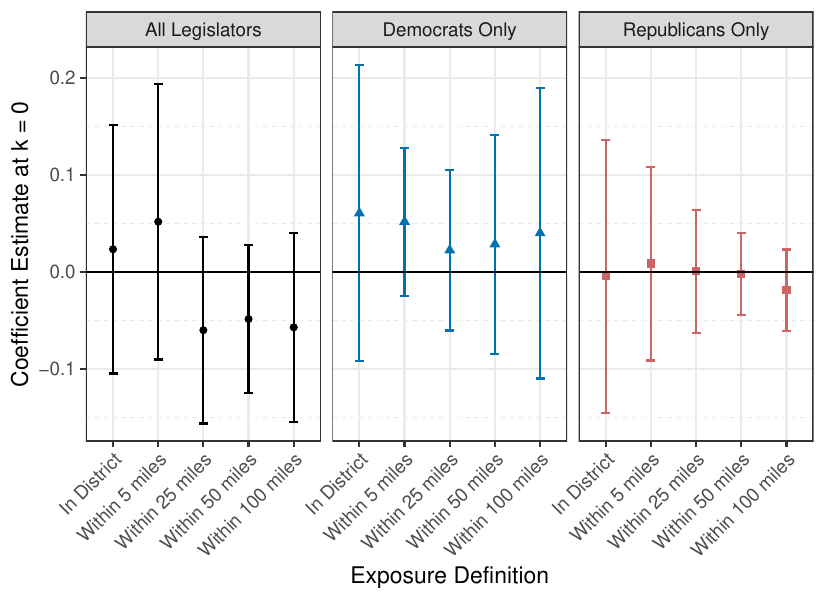}
    \begin{flushleft}
    \scriptsize{\textit{Note}: This figure plots the ATT estimates for the first post-treatment period (\(k = 0\)) pooled across all shootings. Each panel represents a model subset: all treated legislators v. all untreated legislators, treated Democrats v. untreated Democrats, and treated Republicans v. untreated Republicans. For each panel, we also report \(k = 0\) estimates for the broader treatment assignment definitions. 95\% confidence intervals are based on robust standard errors clustered at the state-, legislator-, chamber-, and year levels. Full results are reported in Table \ref{tab:table_main_result}.}    
    \end{flushleft}
    \label{fig:main}
\end{figure}

Figure \ref{fig:main} presents the ATT for the first post-treatment period ($k=0$). To ensure the robustness of our null finding, we report results across 15 distinct specifications, varying both the partisan subset and the geographic definition of ``exposure.'' Our primary specification identifies legislators whose districts contain the shooting location (``In District''), and supplementary models expand this radius to 5, 25, 50, and 100 miles to account for potential geographic spillovers (see Appendix \ref{app:distance}). Following \citet{cengiz_stacked_2019}, we employ a ``first-treatment-only'' approach for multiply-exposed units to avoid bias and preserve power.

The results across all specifications consistently demonstrate that mass shootings do not cause a measurable shift in legislators' Gun Issue Scores, on average. In the full sample (``All Legislators''), the In District exposure estimate is effectively zero ($\mu_{0} = 0.024$, $SE = 0.065$). The 95\% confidence interval ($[-0.105, 0.152]$) is also narrow: when normalized by the standard deviation of latent Gun Issue Scores ($\sigma = 0.732$) across the U.S. for our study period, the interval excludes effects larger than approximately $[-0.14, +0.21]$ standard deviations.

This precisely estimated null persists when the data are disaggregated by party. Neither Democrats nor Republicans show statistically significant or substantively large shifts in their latent scores following a local mass shooting. While we focus on the immediate post-treatment period ($k=0$) to maximize statistical power and minimize attrition bias, the lack of movement across varying geographic radii further suggests that proximity to a mass shooting --- even within one's own constituency --- is insufficient to disrupt the relative equilibrium of polarization.

\subsection{Does Shooting Severity or Context Drive Responsiveness?}
It is also possible that legislative responsiveness is conditional on the specific characteristics of a tragedy. Events with higher fatality counts, or those occurring in particularly sensitive contexts such as schools, might generate sufficient public pressure to overcome partisan constraints. To test this, we estimate pooled ATTs for $k=0$ across several salient subsets of mass shootings: school shootings, workplace incidents, acts of terror, and shootings motivated by hate (a broad category), racism or religious intolerance --- subsets based on perpetrator intentions defined by \citet{violence_project}. We also disaggregate by severity, comparing events with fewer than nine fatalities to those with nine or more, as the average number of victims killed (across all shootings in our study) is approximately 8.5.

\begin{figure}[h!]
    \centering
    \caption{\textbf{ATT Estimates for Effect of Mass Shooting on Gun Issue Score Pooled by Shooting Characteristics}}
    \includegraphics[scale=1]{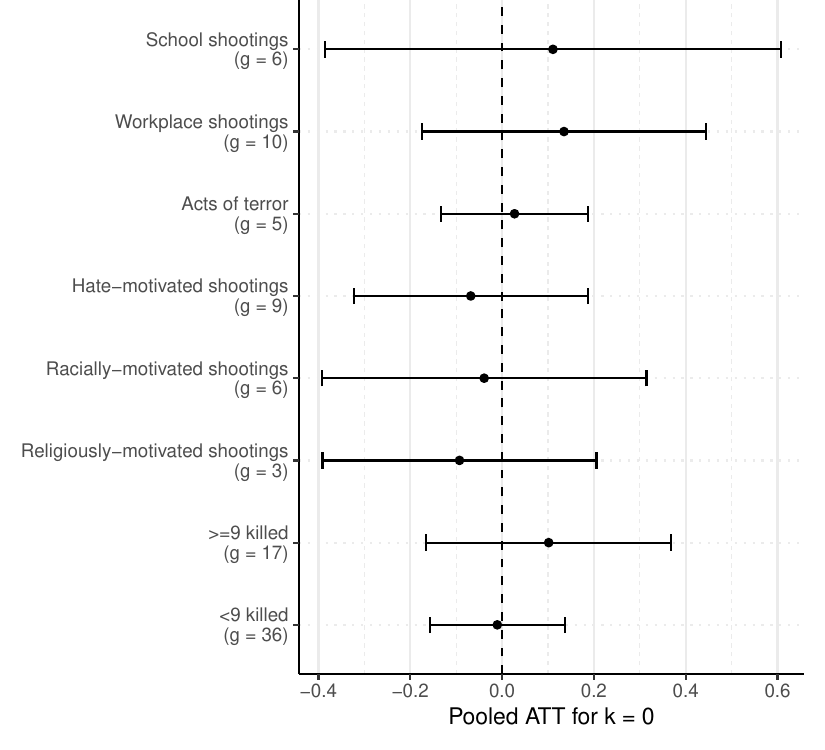}
    \begin{flushleft}
    \scriptsize{\textit{Note}: This figure plots the ATT estimates for the first post-treatment period (\(k = 0\)) pooled across mass shootings according to characteristics identified by \citet{violence_project}. 95\% confidence intervals are based on robust standard errors clustered at the state-, legislator-, chamber-, and year levels. Full results can be seen in Table \ref{tab:table_shooting_type}.}    
    \end{flushleft}
    \label{fig:shooting_characteristics}
\end{figure}

Figure \ref{fig:shooting_characteristics} displays the results of these subset analyses. Notably, not a single estimate across these categories reaches statistical significance. Even in the most acute cases, such as school shootings or high-fatality events, the estimated effect of a mass shooting on legislators' Gun Issue Scores remains negligible and centered near zero. 

These results suggest that the null finding is not driven by the inclusion of mass shootings that fail to generate high levels of public and media salience in our study. Rather, the structural and behavioral constraints of partisan polarization appear robust even to the most heinous acts of violence. Whether a shooting is motivated by racial hatred or targets a school, institutional anchors (be it party discipline or interest-group alignment, for example) prevent individual legislators from translating localized tragedies into a shift in roll-call behavior, on average. This consistent null across shooting characteristics reinforces our conclusion: in the current U.S. political environment, mass shootings serve as powerful focusing events for public discourse, but, more often than not, they stop short of altering the fundamental ideological positions of state representatives.

\section{Robustness}
We conduct several robustness checks to assess the sensitivity of our findings to alternative specifications and measurement choices, with full results reported in the \hyperref[appendix]{Appendix}.

To ensure that our results are not an artifact of our specific measurement strategy, we re-estimate our primary specification using a dynamic IRT model in place of the Gun Issue Score (Appendix \ref{app:irt}). As shown in Figure \ref{fig:irt_did}, this alternative outcome yields substantively identical estimates to our main results. We further test for effects using the proportion of restrictive versus permissive votes taken by a legislator (Appendix \ref{app:alternative_results}) and find no evidence that alternative metrics of legislative activity capture shifts that our latent scores miss.

While our primary analysis estimates the average effect across all shootings to maximize statistical power, Appendix \ref{app:shooting_ests} presents disaggregated, event-specific estimates. These shooting-by-shooting analyses (Figure \ref{fig:disagg_ates}) reveal a degree of heterogeneity that the aggregate effect necessarily masks. While most shootings yield null results, we observe occasional large point estimates in opposing directions. This variation likely reflects idiosyncratic contextual factors, including district characteristics \citep{hassell_mobilize_2020}, victim demographics \citep{markarian2024racially}, and media intensity \citep{Baxter-King2024-fr}. We caution, however, that most event-specific estimates involve only one to three treated legislators. Thus, these ``large'' shifts often reflect high variance in small samples rather than a systematic pattern of responsiveness. Ultimately, these disaggregated results reinforce our conclusion that mass shootings do not generate a consistent, nationally observable shift in legislative behavior.

Appendix \ref{app:sensitivity} reports several sensitivity analyses. In Appendix \ref{app:permutation}, we conduct permutation tests comparing our estimated treatment effects against a null distribution generated by random treatment assignment. As shown in Figure \ref{fig:placebo_violin}, the distribution of shooting-specific estimates is largely consistent with chance, with only 6 of 53 incidents reaching nominal statistical significance. This suggests that the occasional shifts observed in Appendix \ref{app:shooting_ests} are modest relative to the natural session-to-session variation in voting patterns. Additionally, we assess the validity of our identification strategy in Appendix \ref{app:parallel_trends_pre_trends} and find no evidence of systematic pre-treatment divergence that would bias our difference-in-differences estimates.

Returning to the motivating example of the Sandy Hook shooting, we \textit{do} detect a change in Senator John McKinney's Gun Issue Score following the shooting (see Figure \ref{fig:disagg_ates}): in the 2011 session, Sen. McKinney's Gun Issue Score was calculated to be 0.398, and dropped to -0.972 in 2012 and -0.997 by 2013. However, as seen in Figure \ref{fig:placebo_violin}, we cannot conclude that Sen. McKinney's change in voting behavior was greater than what we would expect to observe under random treatment assignment. Representative Mitch Bolinsky is dropped from our analyses as we do not observe him in more than one pre-treatment session.

Finally, we consider whether mass shootings generate electoral pressures that shift a district's representation over time, even if the incumbent's own views remain fixed. Appendix \ref{app:treatment_district} tests for ``replacement effects'' by estimating changes in the Gun Issue Score of \textit{any} legislator representing a treated district, regardless of whether they held office before and after the shooting. Consistent with evidence that public opinion on firearm policy remains relatively stable following shootings \citep{Baxter-King2024-fr, hassell_mobilize_2020}, Figure \ref{fig:app_treat_dist} shows no average shift in Gun Issue Scores among legislators representing affected districts in subsequent sessions.

\section{Conclusion}

In this paper, we introduce the Gun Issue Score, a novel measure of an individual legislator's position on firearm-related policy on a permissive to restrictive scale. By applying this measure to over 15,000 bills and 14,585 legislators across twelve years in all 50 states, we provide detailed insight into how elected officials respond --- or fail to respond --- to acute tragedies within their own constituencies. Our findings qualify canonical theories of focusing events: while mass shootings are profoundly salient tragedies that command immense public and media attention, they do not, on average, disrupt the entrenched partisan polarization that dominates firearm policymaking in the United States.

Across 53 mass shootings and various model specifications, we find that a shooting occurring in (or near) a legislator's district has no detectable effect on their Gun Issue Score. The precision of this null is noteworthy: our 95\% confidence interval excludes shifts larger than approximately $\pm$0.2 standard deviations of the latent Gun Issue Score. This stability persists regardless of the shooting's severity, the context of the incident, or the partisan identity of the representative, on average. 
Even the Sandy Hook shooting --- which, according to media reports prompted two Republican legislators in Connecticut to break ranks and support restrictive legislation --- does not, when examined under our design, constitute a generalizable pattern. Senator McKinney’s voting behavior did shift towards a substantially more restrictive position, but his shift was not unique. In fact, our permutation tests show that shifts of this magnitude occurred as often among untreated legislators. The \textit{appearance} of exceptional responsiveness dissolves against a backdrop of otherwise remarkable stability in session-to-session variation. What looks like a critical juncture in anecdote is, systematically, noise.

Why does this stability prevail? Our mixed-methods design offers some insight. Qualitative interviews with five California legislators --- a state that, notably, exhibits the highest level of partisan polarization on gun policy --- reveal a consistent tension: legislators described personal desires to ``use the moment'' effectively, but also reported powerful constraints and hinted at the weight of party discipline. The admission from one representative that leadership threatened a million-dollar-per-tweet penalty for deviating from the party line illustrates the high stakes of modern polarization on gun policy. In this environment, the focusing event power of a shooting is not lost on the legislator, but its effect may not be observable in their voting behavior --- rather, it is redirected or suppressed by the professional and electoral costs of breaking ranks. These accounts do not prove that such constraints operate uniformly across all states, but they illustrate plausible mechanisms that are consistent with our null finding. In less polarized states or those with weaker party discipline, the balance of forces may differ, and our aggregate estimate averages over that heterogeneity by design.

Our findings thus delineate boundaries of democratic responsiveness in a polarized era. On one hand, the absence of movement following localized tragedies suggests a weakening of the traditional accountability link between district shocks and legislative behavior. On the other, it reflects a political environment in which partisan identity has become the dominant anchor for roll-call voting on firearm policy. So dominant, in fact, that even the most heinous acts of mass violence fail to register a systematic shift in a legislator's behavior, on average. 

It appears that by the time a bill has reached the floor, the calculus of party discipline, interest-group pressure, primary vulnerability, and other explanatory mechanisms have already absorbed the shock of the moment. Given the frequency of these tragedies, perhaps one additional input is rarely the decisive one in the current equilibrium. Unfortunately, we cannot speak to what \textit{would} move legislators off their partisan anchors. What we have shown is that mass shootings, alone, do not appear to be enough.

\endgroup

%---------------------------
\newpage

\begingroup
\setstretch{1}
\bibliographystyle{apsr}
\bibliography{bib} 
\endgroup

\clearpage

\begingroup
\setstretch{1.5}
\clearpage

% \appendix
\setcounter{section}{0}
\setcounter{table}{0}
\setcounter{figure}{0}
\renewcommand{\theHsection}{A.\arabic{section}}
\renewcommand{\thesection}{A.\arabic{section}}
\renewcommand{\thefigure}{A.\arabic{figure}}
\renewcommand{\thetable}{A.\arabic{table}}

\setstretch{1.5}

%----------------------------------
\section*{Appendix}
\label{appendix}

\pagenumbering{arabic}
\setcounter{page}{1}

Intended for online publication only.

\etocdepthtag.toc{mtappendix}
\etocsettagdepth{mtchapter}{none}
\etocsettagdepth{mtappendix}{subsection}
\tableofcontents

\pagebreak

%----------------------------------
\section{Qualitative Interview Procedures}
\label{app:interview}
\renewcommand{\thefigure}{A.1.\arabic{figure}}
\setcounter{figure}{0}
\renewcommand{\thetable}{A.1.\arabic{table}}
\setcounter{table}{0}

To inform our theoretical expectations, we conducted a series of qualitative interviews with California state legislators who represented districts affected by mass shootings that occurred within their district.\footnote{As these interviews were conducted with public officials about their roles as public servants, our project is exempt from IRB regulations typically associated with human subject research. That said, we followed all relevant IRB guidelines to ensure the validity of our findings.} Our sample of 5 interview participants worked out to be a roughly 21\% response rate of all legislators who represented districts affected by mass shootings during our period of study.\footnote{All other affected legislators were contacted, though 5 directly refused to participate. All others either did not respond directly (6 legislators), or stopped responding after we made initial contact (7 legislators). We were unable to make contact with an additional legislator who did not have any contact information publicly available.} All interviews except for the one with Legislator E were conducted on Zoom, and the interview with Legislator E was conducted in their field office, in person, with their chief of staff also present. 

The following questions were used to guide the interview:

\begin{itemize}
    \item How did you first become aware of the mass shooting, and what were your initial reactions upon learning about it?
    \begin{itemize}
        \item Follow-ups:
        \begin{itemize}
            \item Did you feel personally affected by the shooting considering your role as an elected representative for the community?
            \item How did you engage with the affected community members, survivors, and families of victims in the aftermath of the mass shooting?
        \end{itemize}
    \end{itemize}
    \item What were some of the primary concerns expressed by your community following the mass shooting, and how did community leaders address those concerns?
    \begin{itemize}
        \item Follow-ups:
        \begin{itemize}
            \item How did you collaborate with other legislators, government agencies, and community leaders in your response to the mass shooting?
            \item Were there any specific challenges faced during this collaborative effort?
        \end{itemize}
    \end{itemize}
    \item In what ways did the mass shooting in your district influence your legislative priorities and agenda moving forward?
    \begin{itemize}
        \item Follow-ups:
        \begin{itemize}
            \item Can you discuss any legislative measures or policies you proposed or supported in response to the mass shooting? 
            \item What was the rationale behind these proposals?
        \end{itemize}
    \end{itemize}
    \item Were there any specific challenges or obstacles you encountered while attempting to implement legislative or policy changes in response to the mass shooting?
    \begin{itemize}
        \item Follow-ups:
        \begin{itemize}
            \item Did you face any obstacles or support from members of the opposing party? 
            \item Did you face any obstacles or support from interest groups?
        \end{itemize}
    \end{itemize}
    \item Looking back, what do you feel were the most significant lessons learned from your experience responding to the mass shooting, both personally and legislatively?
    \begin{itemize}
        \item Follow-up:
        \begin{itemize}
            \item What advice might you offer to other legislators or leaders whose communities are affected by mass shootings in the future?
        \end{itemize}
    \end{itemize}
    \item Is there anything else that you would like to share with us?
\end{itemize}

While all participants waived anonymity, we ultimately decided to keep all quotations anonymous due to the sensitive nature of the interviews. While all interviews were recorded and transcribed, we will not be publicly sharing the full transcripts of the interviews in order to protect the anonymity of the participants. 

\clearpage

\section{Locations and Details of Mass Shootings}
\label{app:shootings}
\renewcommand{\thefigure}{A.2.\arabic{figure}}
\setcounter{figure}{0}
\renewcommand{\thetable}{A.2.\arabic{table}}
\setcounter{table}{0}

Figure \ref{fig:affected_districts_US} shows a map of affected upper and lower chamber districts, and Table \ref{tab:shootings_in_study} identifies the shootings that are included in our study --- specifying the date, location, number of victims (not including the shooter), and the State Senate and Assembly districts in which the shooting occurred (accounting for redistricting following the 2010 Census). 

\begin{figure}[!h]
    \centering
    \caption{\textbf{U.S. State House and Senate Districts with Mass Shootings, 2011-2022}}
    \includegraphics[scale=0.55]{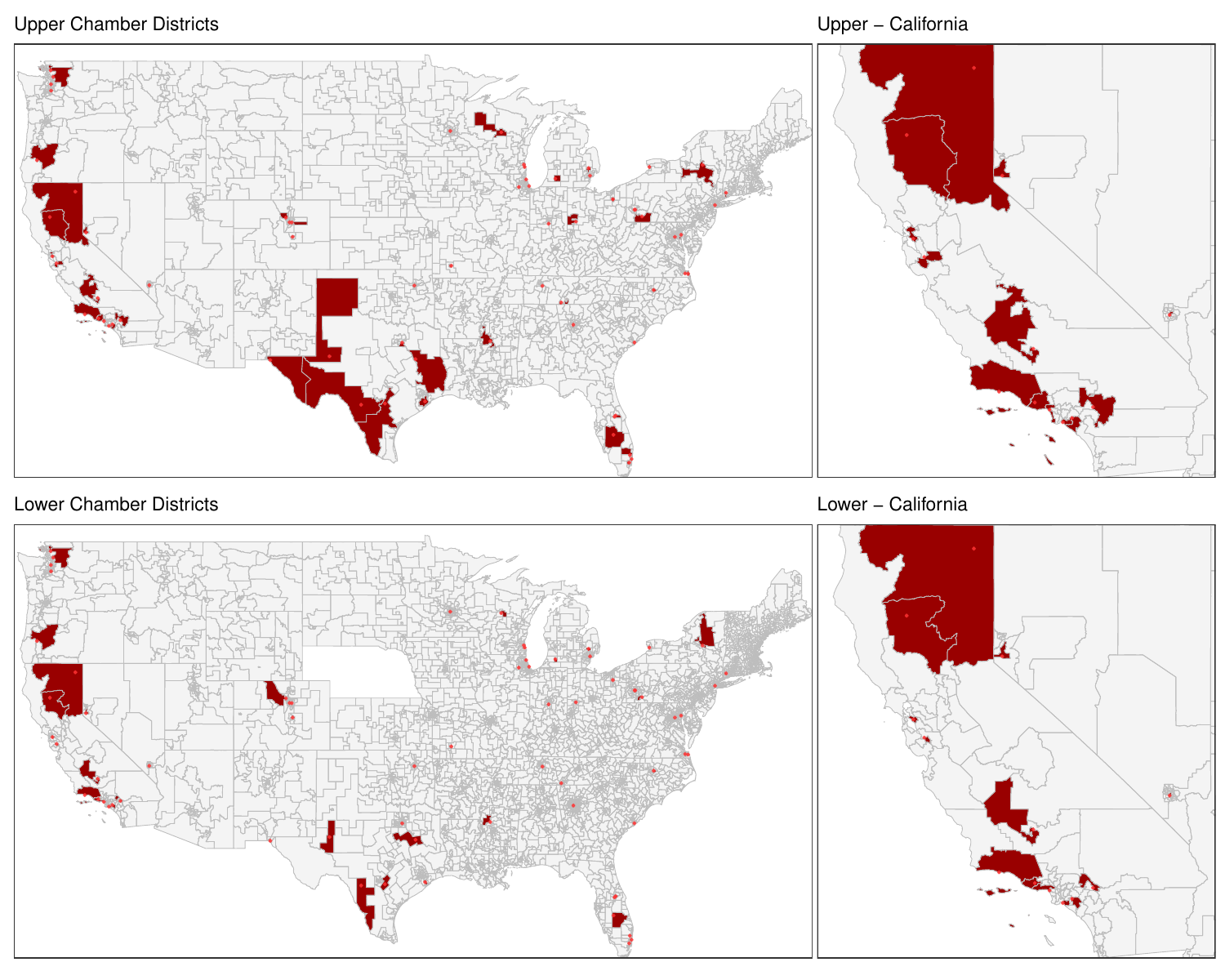}
    \begin{flushleft}
    \scriptsize{\textit{Note}: Red districts are those affected by a mass shooting, shown as the small, light red points at the location of the shooting. The map uses district boundaries adopted after the redistricting cycle following the 2010 Census, and only plots shootings that affected districts under these new maps (accounting for variation in timing of adoption of these maps across states). For a full list of affected districts and shootings (including those happening prior to redistricting), see \ref{app:shootings}. Nebraska has a unicameral state legislature, so we do not report lower chamber districts and only show upper chamber districts on the map.}    
    \end{flushleft}
    \label{fig:affected_districts_US}
\end{figure}

\begin{ThreePartTable}
\fontsize{10}{12}\selectfont

\begin{longtable}{lccccccc}
\caption{\textbf{Mass Shooting Events and Affected Legislative Districts}}
\label{tab:shootings_in_study} \\

\toprule
\toprule
Date & State & City & \# Killed & Lower & Upper & Post-Treat & Included? \\
\midrule
\endfirsthead

\multicolumn{8}{l}{\textbf{Table \ref{tab:shootings_in_study} continued}} \\
\toprule
Date & State & City & \# Killed & Post-Treat & Lower & Upper &  Included? \\
\midrule
\endhead

\midrule
\multicolumn{8}{r}{\textit{Continued on next page}} \\
\endfoot

\bottomrule
\endlastfoot
2010-04-03 & CA & North Hollywood &  4 & 2011 & 42 & 21 & No
\\ 2011-08-07 & OH & Akron &  7 & 2012 & 35 & 28 & No
\\ 2011-09-06 & NV & Carson City &  4 & 2013 & 40 & 16 & No
\\ 2011-10-12 & CA & Seal Beach &  8 & 2012 & 67 & 35 & No
\\ 2012-04-02 & CA & Oakland &  7 & 2012 & 18 & 9 & No
\\ 2012-05-30 & WA & Seattle  &  5 & 2012 & 43 & 43 & No
\\ 2012-07-20 & CO & Aurora & 12 & 2013 & 42 & 29 & Yes
\\ 2012-08-05 & WI & Oak Creek &  6 & 2013 & 21 & 7 & No
\\ 2012-09-27 & MN & Minneapolis  &  6 & 2013 & 59B & 59 & Yes
\\ 2012-12-14 & CT & Sandy Hook & 27 & 2013 & 106 & 28 & Yes
\\ 2013-03-13 & NY & Herkimer &  4 & 2013 & 118 & 51 & Yes
\\ 2013-04-21 & WA & Federal Way  &  4 & 2013 & 30 & 30 & Yes
\\ 2013-06-07 & CA & Santa Monica &  5 & 2013 & 50 & 26 & Yes
\\ 2013-07-26 & FL & Hialeah &  6 & 2014 & 110 & 36 & Yes
\\ 2013-09-16 & DC & Washington D.C. & 12 & 2014 & NA & 6 & No
\\ 2014-02-20 & CA & Alturas &  4 & 2014 & 1 & 1 & Yes
\\ 2014-05-23 & CA & Isla Vista  &  6 & 2014 & 37 & 19 & Yes
\\ 2014-10-24 & WA & Marysville  &  4 & 2015 & 39 & 39 & Yes
\\ 2015-06-17 & SC & Charleston  &  9 & 2015 & 111 & 42 & Yes
\\ 2015-07-16 & TN & Chattanooga &  5 & 2016 & 28 & 10 & Yes
\\ 2015-10-01 & OR & Roseburg  &  9 & 2016 & 7 & 4 & Yes
\\ 2015-11-14 & TX & Tennessee Colony &  6 & 2017 & 8 & 3 & Yes
\\ 2015-12-02 & CA & San Bernardino & 14 & 2016 & 40 & 23 & Yes
\\ 2016-02-20 & MI & Kalamazoo &  6 & 2016 & 61 & 20 & Yes
\\ 2016-03-09 & PA & Wilkinsburg &  5 & 2016 & 24 & 43 & Yes
\\ 2016-06-12 & FL & Orlando  & 49 & 2016 & 47 & 13 & Yes
\\ 2016-07-07 & TX & Dallas &  5 & 2017 & 108 & 23 & Yes
\\ 2016-09-23 & WA & Burlington &  5 & 2017 & 40 & 40 & Yes
\\ 2017-01-06 & FL & Fort Lauderdale &  5 & 2017 & 99 & 34 & Yes
\\ 2017-02-06 & MS & Yazoo City &  4 & 2017 & 51 & 22 & Yes
\\ 2017-03-22 & WI & Rothschild &  4 & 2017 & 85 & 29 & Yes
\\ 2017-06-05 & FL & Orlando  &  5 & 2017 & 49 & 13 & Yes
\\ 2017-10-01 & NV & Las Vegas  & 60 & 2019 & 16 & 10 & Yes
\\ 2017-11-05 & TX & Sutherland Springs & 25 & 2019 & 44 & 21 & Yes
\\ 2017-11-14 & CA & Rancho Tehama Reserve &  5 & 2018 & 3 & 4 & Yes
\\ 2018-01-28 & PA & Melcroft, Saltlick Township &  4 & 2018 & 52 & 32 & Yes
\\ 2018-02-14 & FL & Parkland & 17 & 2018 & 96 & 29 & Yes
\\ 2018-02-26 & MI & Detroit &  4 & 2018 & 7 & 4 & Yes
\\ 2018-04-22 & TN & Antioch &  4 & 2018 & 52 & 21 & Yes
\\ 2018-05-18 & TX & Santa Fe & 10 & 2019 & 24 & 11 & Yes
\\ 2018-06-28 & MD & Annapolis &  5 & 2018 & 30A & 30 & Yes
\\ 2018-09-12 & CA & Bakersfield &  5 & 2019 & 32 & 14 & Yes
\\ 2018-10-27 & PA & Pittsburgh & 11 & 2019 & 23 & 43 & Yes
\\ 2018-11-07 & CA & Thousand Oaks & 12 & 2019 & 44 & 27 & Yes
\\ 2019-01-23 & FL & Sebring  &  5 & 2019 & 55 & 26 & Yes
\\ 2019-02-15 & IL & Aurora &  5 & 2019 & 83 & 42 & Yes
\\ 2019-05-31 & VA & Virginia Beach & 12 & 2019 & 84 & 8 & Yes
\\ 2019-08-03 & TX & El Paso & 23 & 2021 & 76 & 29 & Yes
\\ 2019-08-04 & OH & Dayton &  9 & 2020 & 39 & 5 & Yes
\\ 2019-08-31 & TX & Midland-Odessa &  7 & 2021 & 82 & 31 & Yes
\\ 2019-12-10 & NJ & Jersey City &  4 & 2020 & 31 & 31 & Yes
\\ 2020-02-26 & WI & Milwaukee &  5 & 2020 & 18 & 6 & Yes
\\ 2020-03-15 & MO & Springfield &  4 & 2020 & 135 & 30 & Yes
\\ 2021-01-09 & IL & Chicago &  5 & 2021 & 25 & 13 & Yes
\\ 2021-03-16 & GA & Atlanta &  8 & 2021 & 57 & 36 & Yes
\\ 2021-03-22 & CO & Boulder & 10 & 2021 & 13 & 18 & Yes
\\ 2021-03-31 & CA & Orange &  4 & 2021 & 68 & 37 & Yes
\\ 2021-04-15 & IN & Indianapolis  &  8 & 2021 & 93 & 36 & Yes
\\ 2021-05-26 & CA & San Jose &  9 & 2021 & 27 & 15 & Yes
\\ 2021-11-30 & MI & Oxford &  4 & 2022 & 46 & 12 & Yes
\\ 2021-12-27 & CO & Denver &  5 & 2022 & 2 & 32 & Yes
\\ 2022-05-14 & NY & Buffalo  & 10 & 2022 & 141 & 63 & Yes
\\ 2022-05-24 & TX & Uvalde & 21 & 2023 & 80 & 19 & No
\\ 2022-06-01 & OK & Tulsa &  4 & 2022 & 79 & 39 & No
\\ 2022-07-04 & IL & Highland Park &  7 & 2023 & 58 & 29 & No
\\ 2022-10-13 & NC & Raleigh &  5 & 2023 & 39 & 14 & No
\\ 2022-11-19 & CO & Colorado Springs &  5 & 2023 & 16 & 10 & No
\\ 2022-11-22 & VA & Chesapeake &  6 & 2023 & 78 & 5 & No
\\ \bottomrule
%     \\[-1.8ex]\hline 
%     \hline \\[-1.8ex]
% \insertTableNotes

\end{longtable}
\vspace{-2.2ex}

\begin{tablenotes}[flushleft]
\setlength\labelsep{0pt}
\scriptsize
\item \textit{Note:} Data on shooting location and number of victims (excluding the shooter) come from the Violence Project. Upper and Lower refer to the upper- and lower-chamber districts in which the shooting occurred, identified by mapping shooting coordinates to state legislative district shapefiles from the U.S. Census Bureau and cross-referenced with the Gun Violence Archive when available. The 2010 North Hollywood shooting is excluded due to lack of pre-treatment observations. Post-Treat reports the year corresponding to \(k = 0\), defined as the shooting year if the event occurs in the first six months of the year, and the subsequent year otherwise. For shootings predating post-2010 redistricting, districts are reported under pre-Census maps and cross-referenced to ensure continued treatment assignment.
\end{tablenotes}
\vspace{2ex}

\end{ThreePartTable}

Table \ref{tab:shootings_in_study} highlights an important feature of the data worth considering: these mass shootings, like many others across the United States, are concentrated in urban areas with higher rates of population density \citeApp{violence_project, kwon_mass_shootings_2019}.

When we consider a broader definition of shootings and expand our focus to include all known/recorded incidents of gun violence in the United States, we find that there is effectively no variation in treatment assignment. Focusing on mass shootings therefore balances theoretical relevance with the empirical requirement for identifiable counterfactuals, reinforcing the validity of our design.

\begin{figure}[!h]
    \centering
    \caption{\textbf{U.S. State House and Senate Districts with Any Shooting, 2019-2025}}
    \includegraphics[scale=.55]{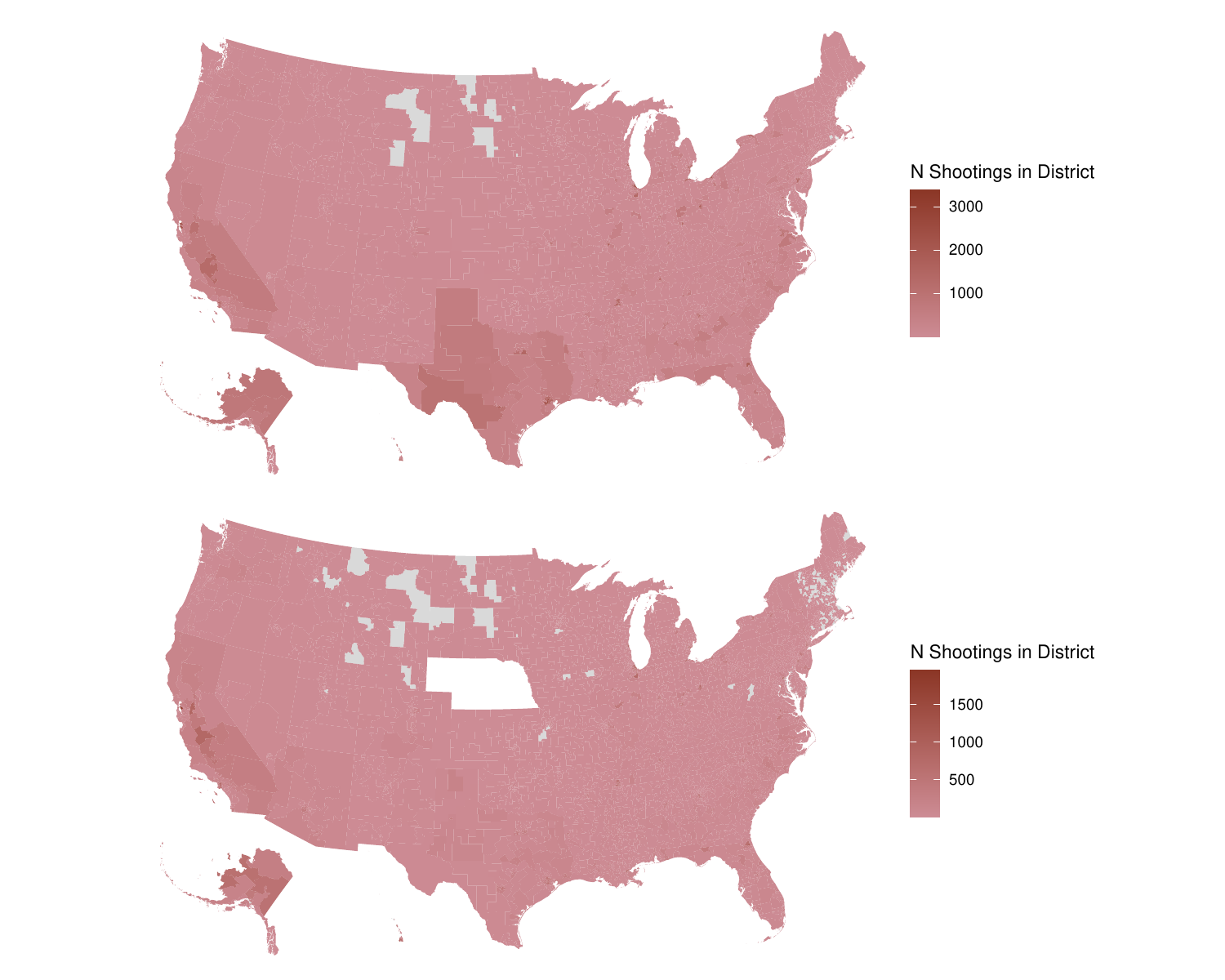}
    \begin{flushleft}
    \scriptsize{\textit{Note}: The top panel shows Upper Chamber Legislative Districts, while the bottom panel shows Lower Chamber Legislative Districts. 
    1,935/1,949 (99\%) of Upper Districts had at least 1 shooting between 01/01/2019 - 01/01/2025, while 4,573/4,790 (95\%) of Lower Districts had at least 1 shooting. District shading reflects the total number of shootings recorded within legislative district boundaries between January 1, 2019 and January 1, 2025, using data from the Gun Violence Archive. Districts with no recorded shootings are shaded in light gray. Alaska and Hawaii are shown as insets for visual clarity. Legislative district boundaries are from the U.S. Census Bureau’s 2019 State Legislative District shapefiles.} 
    \end{flushleft}
    \label{fig:gva_map}
\end{figure}

\clearpage

\section{Validating the GLC Legislative Scoring Regime}
\label{app:GLC}
\renewcommand{\thefigure}{A.3.\arabic{figure}}
\setcounter{figure}{0}
\renewcommand{\thetable}{A.3.\arabic{table}}
\setcounter{table}{0}

As the GLC claims on their website, ``Giffords is an organization dedicated to saving lives from gun violence.'' In other official communications, the GLC states that it ``is fighting to end the gun lobby’s stranglehold on our political system,'' which may reasonably lead one to believe that the GLC scoring system could suffer from pro-gun control bias in their scoring of state firearm-related legislation. To validate their scoring regime, we hired a team of research assistants (RAs) to hand-code a sample of the bills also coded by the GLC to assess the degree to which the GLC data may have been biased due to the original coder's association with a gun control advocacy organization. We provided the RAs with instruction on the types of firearms policy that exist within the United States and instructed them to code bills on a -1 to +1 scale, with -1 being considered a ``permissive'' bill that makes it easier for individuals to access/purchase/own/possess firearms, 0 being a neutral policy, and +1 being a ``restrictive'' policy that makes it harder for someone to  access/purchase/own/possess firearms. 

After coding samples of bills from all states and from California in particular, we found that our RAs had an average of a 73.5\% match with the GLC coding scheme, and 80\% consistency with each other. After digging into this further, we determined that the majority of mismatches between the RAs and the GLC were from the RAs considering many of the procedural budgetary bills as neutral, and the GLC coding some bills our RAs coded as neutral as ``strengthen'' or ``weaken'' based on very specific clauses or items in the bills that our RAs had not considered to be as relevant to the overall bill. Overall, we are confident that our RAs' coding is consistent with the GLC coding scheme, and given the small number of uncoded bills in California (25), we used the bill ranking provided by the GLC when available, and filled in all missing bill rankings with assessments from our RAs (and confirmed by the authors). We are therefore left with 366 bills that were voted on between 2011 and 2022 in the California Legislature.

\begin{table}[!ht]
    \centering
    \scriptsize
    \caption{\textbf{Examples of Firearm-Related Legislation and Coding in California}}
    \label{tab:bill_examples}
    \begin{threeparttable}
    \begin{tabular}{@{} llp{1.5cm}p{0.28\linewidth}cc @{}}
    \toprule
    \toprule
    Session Year & Bill Number & Status & Description & GLC Ranking & Bill Effect \\
    \midrule
    %2011 & S.B.610 & Enacted & Concealed Carry Weapons applicants are exempt from paying for required training courses or obtaining liability insurance. Licensing authorities must notify applicants of approval or denial within 90 days, providing specific reasons for any denial. & Weaken & --1 \\
    %\midrule
    2013 & A.B.1014 & Enacted & Allows family members or law enforcement to petition for a Gun Violence Restraining Order (GVRO) if there is evidence that an individual poses a danger. The GVRO temporarily prohibits the individual from purchasing or possessing firearms or ammunition and permits law enforcement to remove existing firearms or ammunition. & Strengthen & +1 \\
    \midrule
    2013 & S.B.916 & Failed & This bill would allow a handgun manufacturer to return a model to the CA handgun roster without retesting if it was removed for reasons other than failing testing. It would expand exemptions for ``new model'' designation to include minor changes and permit dealers to sell off-roster handguns within 30 days. & Weaken & --1 \\
    \midrule
    2019 & A.B.1009 & Vetoed & This bill allows firearm transaction records to be submitted electronically, instead of by mail or in person, and authorizes the CA DOJ to charge reasonable processing fees for forms submitted by mail or in person. & Neutral & 0 \\
    %\midrule
    %2021 & A.B.178 & Enacted & This Budget bill allocates \$40 million for a court-based firearm relinquishment program and an additional \$3 million to the UC Firearm Violence Research Center. & Strengthen & +1 \\    
    \bottomrule
    \bottomrule
    \end{tabular}
    \begin{tablenotes}[flushleft]
\setlength\labelsep{0pt}
\scriptsize
\item \textit{Note:} Bill descriptions and coding provided conditionally as a part of a data sharing agreement with the Giffords Law Center (GLC) and cannot be shared as a part of the replication materials. Descriptions of bills have been edited from original bill descriptions provided by the GLC to remove any confidential information.
\end{tablenotes}
\end{threeparttable}

\end{table}

\clearpage

\section{Treatment Timing}
\label{app:treat_timing}
\renewcommand{\thefigure}{A.4.\arabic{figure}}
\setcounter{figure}{0}
\renewcommand{\thetable}{A.4.\arabic{table}}
\setcounter{table}{0}

\begin{figure}[h!]
    \centering
    \caption{\textbf{Treatment Timing for Legislators Representing Districts in Which a Mass Shooting Occurred}}
    \label{fig:treatment_timing}
    \includegraphics[scale=.45]{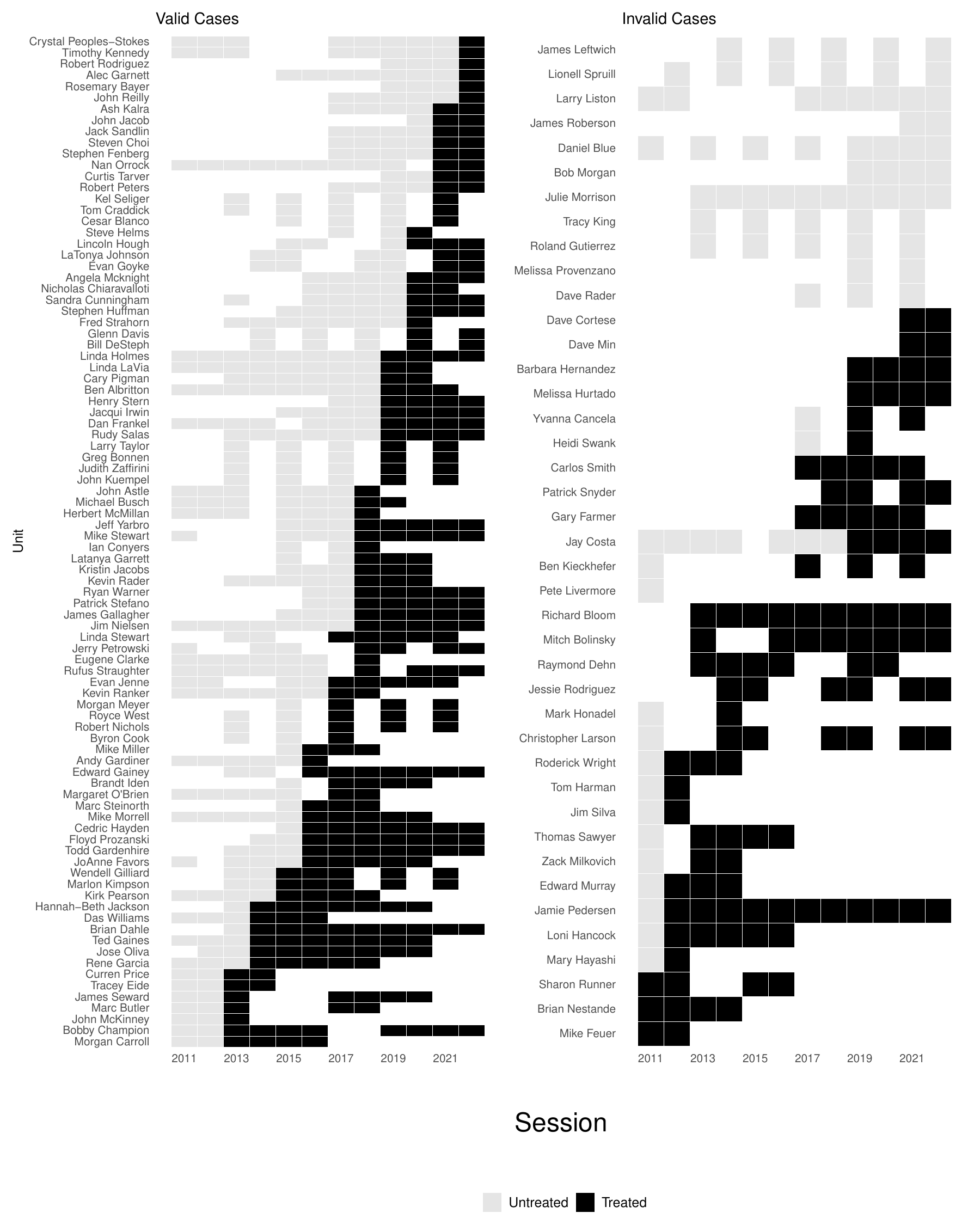}
    \begin{flushleft}
    \scriptsize{\textit{Note}: Retained observations are shown on the left-most panel, and dropped observations appear on the right-most panel. A shooting in months 1-6 of a year leads to that year being considered post-treatment, while a shooting occurring in months 7-12 of a given year leads to that year being considered pre-treatment. Units are dropped if a legislator is treated by multiple shootings (only the case for Jay Costa of Pennsylvania), or if the legislator does not have at least two pre-shooting and one post-shooting observation.}
    \end{flushleft}
\end{figure}

\clearpage

\section{Treatment at Varying Distances}
\label{app:distance}
\renewcommand{\thefigure}{A.5.\arabic{figure}}
\setcounter{figure}{0}
\renewcommand{\thetable}{A.5.\arabic{table}}
\setcounter{table}{0}

We rely on state legislative district shapefiles from the United States Census Bureau\footnote{\url{https://www.census.gov/geographies/mapping-files.html}.} to identify district boundaries following the 2012 redistricting cycle. These boundaries are used for shootings occurring after the adoption of the new district boundaries, which varies across states in our study. For shootings occurring before the redistricting done following the 2010 Census, we report the district that was affected under the earlier maps, and cross-reference district boundaries after the Census to ensure that legislators who were redistricted to represent a new area are still considered treated if they previously represented an affected district. District shapefiles from this earlier period are sourced primarily from the Redistricting Data Hub.\footnote{\url{https://redistrictingdatahub.org}.} Shooting locations are from the Violence Project,\footnote{\url{https://www.theviolenceproject.org}.} and are cross-referenced with records from the Gun Violence Archive,\footnote{\url{https://www.gunviolencearchive.org}.} when available. 

To determine shootings that are affected under the varying distance, we calculate a 5 and 25 mile radii, in euclidean distance, around each shooting's location. 
We use the North American Datum of 1983 (NAD83) geodetic datum as our Coordinate Reference System (CRS).
Figure \ref{fig:dist_ca} shows these buffers, in blue and green, respectively. Figure \ref{fig:dist_ca} zooms in on Southern California and shows shootings occurring between 2012 and 2022 only.

\begin{figure}[!h]
    \centering
    \caption{\textbf{Treatment Assignment Strategy for Varying Distances}}
    \label{fig:dist_ca}
    \includegraphics[scale=0.8]{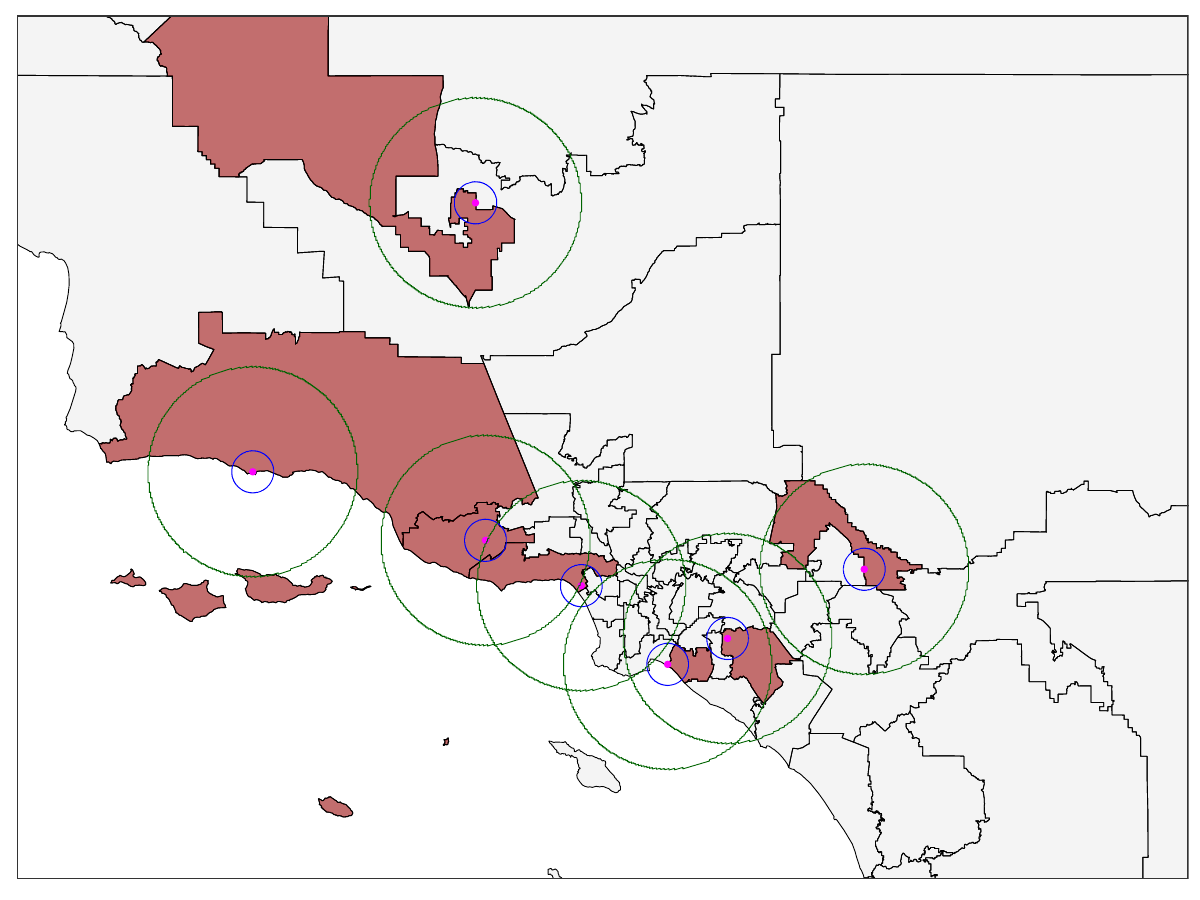}
    \begin{flushleft}
    \scriptsize{\textit{Note}: Legislative district shapefiles used to determine shooting location and affected districts vary across states and redistricting cycles. In this figure, we use boundaries defined after the 2012 redistricting in California, and zoom in on shootings occurring in Southern California between 2012 and 2022. We only map lower chamber districts in this figure as an example. Districts shaded red as those in which the shooting (red points) occurred. The blue buffer represents a 5 mile radius (euclidean distance) from the shooting's location, while green buffers are 25 mile radii.}    
    \end{flushleft}
\end{figure}

\clearpage

\section{Full Reporting of Main Results}
\label{app:main_extended}
\renewcommand{\thefigure}{A.6.\arabic{figure}}
\setcounter{figure}{0}
\renewcommand{\thetable}{A.6.\arabic{table}}
\setcounter{table}{0}

\subsection{Main Model}
\label{app:result_1}
\begin{table}[!htp]
\centering
\scriptsize
\caption{\textbf{Full Reporting of Results from Figure \ref{fig:main}}}
\label{tab:table_main_result}
\begin{threeparttable}
\begin{tabular}{@{\extracolsep{0pt}}lccccc}
\toprule
\toprule
 & \multicolumn{5}{c}{ATT at $k=0$} \\
\cmidrule(lr){2-6}
Specification & In District & Within 5 mi & Within 25 mi & Within 50 mi & Within 100 mi\\
\midrule
All Legislators &\makecell{0.024 \\ (0.065)} & \makecell{0.052 \\ (0.073)} & \makecell{-0.060 \\ (0.049)} & \makecell{-0.048 \\ (0.039)} & \makecell{-0.057 \\ (0.050)} \\
\\  & \makecell{90 treated of \\4208 legislators} & \makecell{408 treated of \\4186 legislators} & \makecell{1164 treated of \\3859 legislators} & \makecell{1524 treated of \\3608 legislators} & \makecell{1812 treated of \\3274 legislators} \\
\midrule
\\ Democrats Only &\makecell{0.061 \\ (0.078)} & \makecell{0.052 \\ (0.039)} & \makecell{0.023 \\ (0.042)} & \makecell{0.029 \\ (0.058)} & \makecell{0.040 \\ (0.077)} \\
\\  & \makecell{48 treated of \\2035 legislators} & \makecell{263 treated of \\2017 legislators} & \makecell{637 treated of \\1810 legislators} & \makecell{767 treated of \\1685 legislators} & \makecell{896 treated of \\1577 legislators} \\
\midrule
\\ Republicans Only &\makecell{-0.004 \\ (0.072)} & \makecell{0.009 \\ (0.051)} & \makecell{0.001 \\ (0.032)} & \makecell{-0.002 \\ (0.022)} & \makecell{-0.019 \\ (0.021)} \\
\\  & \makecell{42 treated of \\2178 legislators} & \makecell{145 treated of \\2174 legislators} & \makecell{528 treated of \\2054 legislators} & \makecell{758 treated of \\1928 legislators} & \makecell{918 treated of \\1701 legislators} \\
\bottomrule
\bottomrule
\end{tabular}

\begin{tablenotes}[flushleft]
\scriptsize
\item \textit{Note}: Robust standard errors are clustered at the state-, legislator-, chamber- and year-levels and reported in parentheses. Table reports the estimated effects of mass shootings occurring within a legislator's district on a legislator's Gun Issue Score for the first post-treatment period (\(k = 0\)).
\end{tablenotes}

\end{threeparttable}
\end{table}

\clearpage

\subsection{State Heterogeneity}
\label{app:result_2}

Table \ref{tab:table_state_heterogeneity} presents disaggregated $k=0$ estimates for the In District model by state to explore this potential heterogeneity in responsiveness. The column "All Legislators" compares all treated units to untreated units, while the subsequent columns report results within partisan subsets. Empty cells indicate that shootings may not have occurred in a district represented by a legislator of a given party. 

First, we find evidence of partisan reinforcement, where mass shootings cause legislators to double down on existing partisan positions rather than moderating or becoming more aligned with their out-party. 
In Michigan, Republicans shifted sharply toward a more permissive stance ($\mu_{0, R} = 1.216$), with similar movements occurring among Republicans in Tennessee ($\mu_{0, R} = 0.271$), and to a lesser extent in Pennsylvania ($\mu_{0, R} = 0.038$). 
Likewise, Democrats in Georgia became more restrictive post-shooting ($\mu_{0, D} = -0.369$), as did Democrats in NY ($\mu_{0, D} = -0.280$) and OH ($\mu_{0, D} = -0.169$). 
In these contexts, legislators do appear to be ``responsive'' to mass shootings, though the action that the affected representatives take is to signal stronger alignment with their party’s typical firearm policy platform.

Second, we observe cases where treated legislators align more closely with the state's dominant political equilibrium, potentially reflecting the pressures of being in a legislative minority or a highly polarized environment. Affected Democrats in Minnesota actually became more permissive relative to their co-partisans ($\mu_{0, D} = 0.419$), while Republicans in California ($\mu_{0, R} = -0.211$) and Connecticut ($\mu_{0, R} = -1.408$) became significantly more restrictive. The latter estimate for Connecticut Republicans represents the single largest estimate from any of our models, following the Sandy Hook Elementary School shooting.

Finally, in states like Washington ($g = 3$), we observe partisan cross-pressures that result in aggregate nulls. Here, both parties moved away from their traditional poles: treated Democrats shifted toward more permissive positions ($\mu_{0, D} = 0.295$) relative to untreated co-partisans, while treated Republicans shifted toward more restrictive positions ($\mu_{0, R} = -0.050$). These diverging marginal changes effectively cancel each other out in the aggregate, even if they capture meaningful shifts among affected lawmakers.
\begin{table}[!htp]
\centering
\scriptsize
\caption{\textbf{Heterogeneity in State-Level Estimates}}
\label{tab:table_state_heterogeneity}
\begin{threeparttable}
\begin{tabular}{@{\extracolsep{0pt}}lccc}
\toprule
\toprule
 & \multicolumn{3}{c}{Gun Issue Score} \\
\cmidrule(lr){2-4}
State & All Legislators & Democrats & Republicans \\
\midrule
CA (g = 9) & \makecell{-0.029 \\ (0.160)} & \makecell{0.099 \\ (0.256)} & \makecell{-0.211 \\ (0.069)} \\
\midrule
CO (g = 3) & \makecell{0.354 \\ (0.335)} & \makecell{0.232 \\ (0.243)} &  \\
\midrule
CT (g = 1) & \makecell{-0.900 \\ (0.062)} &  & \makecell{-1.408 \\ (0.072)} \\
\midrule
FL (g = 6) & \makecell{-0.234 \\ (0.154)} & \makecell{-0.130 \\ (0.123)} & \makecell{-0.023 \\ (0.064)} \\
\midrule
GA (g = 1) & \makecell{-1.030 \\ (0.045)} & \makecell{-0.369 \\ (0.025)} &  \\
\midrule
IL (g = 2) & \makecell{0.035 \\ (0.342)} & \makecell{0.161 \\ (0.379)} &  \\
\midrule
IN (g = 1) & \makecell{0.472 \\ (0.606)} &  & \makecell{0.419 \\ (0.575)} \\
\midrule
MD (g = 1) & \makecell{-0.274 \\ (0.050)} & \makecell{-0.290 \\ (0.170)} & \makecell{-0.227 \\ (0.191)} \\
\midrule
MI (g = 3) & \makecell{0.507 \\ (0.290)} & \makecell{0.410 \\ (0.573)} & \makecell{1.216 \\ (0.246)} \\
\midrule
MN (g = 1) & \makecell{0.619 \\ (0.141)} & \makecell{0.419 \\ (0.200)} &  \\
\midrule
MO (g = 1) & \makecell{-0.200 \\ (0.278)} &  & \makecell{0.072 \\ (0.211)} \\
\midrule
NJ (g = 1) & \makecell{0.042 \\ (0.032)} & \makecell{-0.069 \\ (0.069)} &  \\
\midrule
NY (g = 2) & \makecell{-0.462 \\ (0.147)} & \makecell{-0.280 \\ (0.061)} & \makecell{-0.053 \\ (0.113)} \\
\midrule
OH (g = 1) & \makecell{-0.096 \\ (0.048)} & \makecell{-0.169 \\ (0.025)} & \makecell{-0.009 \\ (0.057)} \\
\midrule
OR (g = 1) & \makecell{0.323 \\ (0.183)} & \makecell{0.316 \\ (0.173)} & \makecell{0.187 \\ (0.266)} \\
\midrule
PA (g = 3) & \makecell{0.016 \\ (0.053)} & \makecell{-0.034 \\ (0.032)} & \makecell{0.038 \\ (0.011)} \\
\midrule
SC (g = 1) & \makecell{0.035 \\ (0.192)} & \makecell{0.013 \\ (0.120)} &  \\
\midrule
TN (g = 2) & \makecell{0.087 \\ (0.055)} & \makecell{-0.112 \\ (0.151)} & \makecell{0.271 \\ (0.124)} \\
\midrule
TX (g = 6) & \makecell{0.275 \\ (0.228)} & \makecell{0.727 \\ (0.399)} & \makecell{0.138 \\ (0.115)} \\
\midrule
WA (g = 3) & \makecell{0.021 \\ (0.044)} & \makecell{0.295 \\ (0.115)} & \makecell{-0.050 \\ (0.012)} \\
\bottomrule
\bottomrule
\end{tabular}

\begin{tablenotes}[flushleft]
\scriptsize
\item \textit{Note}: Robust standard errors are clustered at the legislator, chamber, and year levels and reported in parentheses. Table reports the estimated effects of mass shootings occurring within a legislator's district on a legislator's Gun Issue Score for the first post-treatment period (\(k = 0\)). $g$ refers to the number of shootings within each state for the model encompassing all legislators, regardless of party. 
\end{tablenotes}

\end{threeparttable}
\end{table}

\clearpage

\subsection{Shooting Characteristics}
\label{app:result_3}
\begin{table}[!htp]
\centering
\scriptsize
\caption{\textbf{Full Reporting of Results from Figure \ref{fig:shooting_characteristics}}}
\label{tab:table_shooting_type}
\begin{threeparttable}
\begin{tabular}{@{\extracolsep{0pt}}lccc}
\toprule
\toprule
 & \multicolumn{3}{c}{ATT at $k=0$} \\
\cmidrule(lr){2-4}
Shooting Characteristic & All Legislators & Democrats Only & Republicans Only \\
\midrule
School Shooting & \makecell{0.111 \\ (0.254)} & \makecell{-0.148 \\ (0.224)} & \makecell{0.151 \\ (0.203)} \\
$g=$ 6 & \makecell{11 treated of \\799 legislators} & \makecell{4 treated of \\390 legislators} & \makecell{7 treated of \\410 legislators} \\
\midrule
Workplace Shooting & \makecell{0.135 \\ (0.158)} & \makecell{-0.047 \\ (0.240)} & \makecell{-0.018 \\ (0.186)} \\
$g=$ 10 & \makecell{16 treated of \\881 legislators} & \makecell{9 treated of \\434 legislators} & \makecell{7 treated of \\448 legislators} \\
\midrule
Acts of Terror & \makecell{0.027 \\ (0.082)} & \makecell{-0.205 \\ (0.143)} & \makecell{0.062 \\ (0.241)} \\
$g=$ 5 & \makecell{11 treated of \\674 legislators} & \makecell{5 treated of \\309 legislators} & \makecell{6 treated of \\367 legislators} \\
\midrule
Hate-motivated Shooting & \makecell{-0.068 \\ (0.130)} & \makecell{-0.077 \\ (0.087)} & \makecell{-0.198 \\ (0.022)} \\
$g=$ 9& \makecell{16 treated of \\1367 legislators} & \makecell{13 treated of \\682 legislators} & \makecell{3 treated of \\688 legislators} \\
\midrule
Racially-motivated Shooting & \makecell{-0.039 \\ (0.180)} & \makecell{-0.064 \\ (0.105)} & \\
$g=$ 6& \makecell{10 treated of \\1045 legislators} & \makecell{10 treated of \\528 legislators} & \\
\midrule
Religiously-motivated Shooting & \makecell{-0.093 \\ (0.152)} & \makecell{-0.132 \\ (0.084)} & \\ 
$g=$ 3 & \makecell{6 treated of \\535 legislators} & \makecell{6 treated of \\316 legislators} & \\
\midrule
$>=9$ Killed & \makecell{0.102 \\ (0.136)} & \makecell{0.107 \\ (0.141)} & \makecell{-0.162 \\ (0.163)} \\
$g=$ 17 & \makecell{28 treated of \\1797 legislators} & \makecell{16 treated of \\928 legislators} & \makecell{12 treated of \\871 legislators} \\
\midrule
$<9$ Killed & \makecell{-0.010 \\ (0.075)} & \makecell{0.043 \\ (0.093)} & \makecell{0.059 \\ (0.076)} \\
$g=$ 36 & \makecell{62 treated of \\3258 legislators} & \makecell{32 treated of \\1530 legislators} & \makecell{30 treated of \\1733 legislators} \\
\bottomrule
\bottomrule
\end{tabular}

\begin{tablenotes}[flushleft]
\scriptsize
\item \textit{Note}: Robust standard errors are clustered at the state-, legislator-, chamber- and year-levels and reported in parentheses. Table reports the estimated effects of mass shootings occurring within a legislator's district on a legislator's Gun Issue Score for the first post-treatment period (\(k = 0\)). $g$ refers to the number of shootings within each model encompassing all legislators, regardless of party. 
\end{tablenotes}

\end{threeparttable}
\end{table}

\clearpage

\section{Bayesian IRT}
\label{app:irt}
\renewcommand{\thefigure}{A.7.\arabic{figure}}
\setcounter{figure}{0}
\renewcommand{\thetable}{A.7.\arabic{table}}
\setcounter{table}{0}

Leading scholars in political methodology have developed various methods to estimate legislators' ideal points using data from roll-call votes, survey responses, campaign contributions, political speeches, and social media. But how do these ideological measures translate to a context in which we wish to specifically capture a legislator's positionality on a policy domain as particular as gun control? Considering that other scholars argue that gun control preferences are not effectively computed through ideology estimates derived from a legislator's \textit{overall} roll call voting behavior \citeApp{ansolabehere_roll_call_2001}, we are constrained in our ability to rely on such canonical measures.

We choose to employ the Bradley–Terry model to pairwise compare the legislators in terms of supporting restrictive gun policies \citeApp{bradley1952rank}. Similar to the DW-NOMINATE scores, our scores reveal the \textit{relative} positions of legislators on gun control in each legislative session. In terms of measuring state legislators' stances on specific issue areas, two other common practices are to 1) use survey responses from the individual legislators, and 2) use the spatial choice model or Bayesian Item Response Theory (IRT) model on a subset of roll-call voting records \citeApp{jeong2018measuring}. We discuss these alternative measures of gun-policy-specific ideal points in Sections \ref{app:NRAscores} and \ref{app:IRTscores}, and estimate the effects of shooting using these alternative measures in \ref{app:alternative_results}. 

\clearpage

\subsection{Bayesian IRT Scores}
\label{app:IRTscores}
We also estimate the latent positions of state legislators on firearm-related issues using the Bayesian IRT model (implemented through the \texttt{brms} package in R) and the same subset of roll-call votes as described in Section \ref{score_method} \citeApp{clinton2004statistical, bafumi2005practical}. Figure \ref{fig:irt_comparison} compares the Gun Issue Score and the estimated gun-related ideal points. 

\begin{figure}[!h]
    \centering
    \caption{\textbf{Bayesian IRT Estimates Compared to Gun Issue Score}}
    \includegraphics[scale=1]{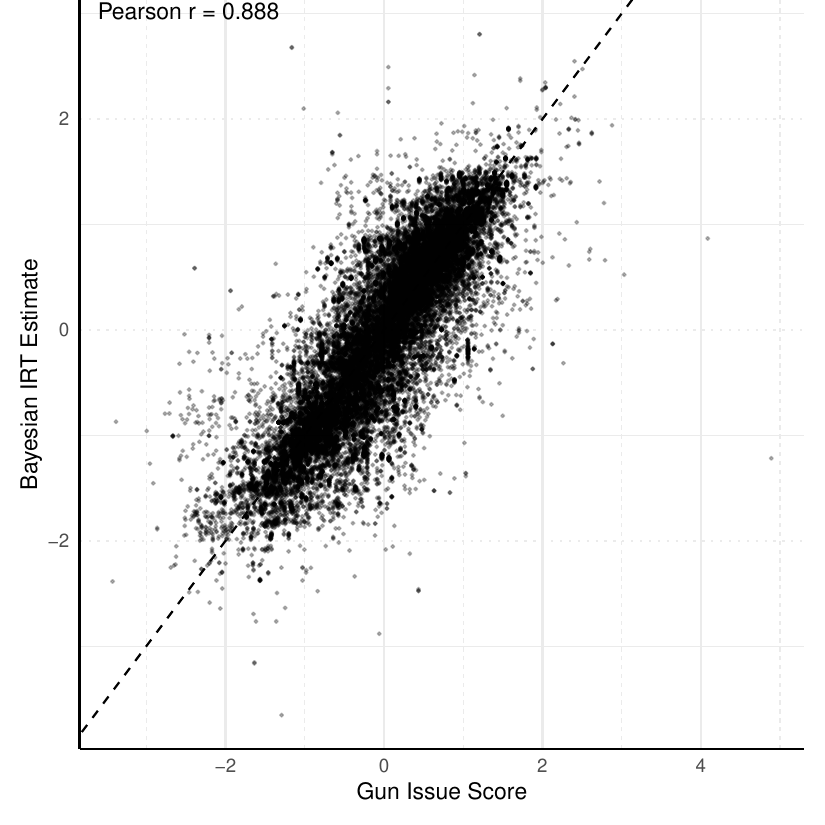}
    \begin{flushleft}
    \scriptsize{\textit{Note}: This figure plots the correlation between the Bayesian IRT estimates and the Gun Issue Score.}
    \end{flushleft}
    \label{fig:irt_comparison}
\end{figure}

\clearpage

\subsection{Results Using IRT Estimates}
\label{app:IRTresults}

\begin{figure}[h!]
    \centering
    \caption{\textbf{ATT Estimates for Staggered DiD Demonstrate Null Results Across Party When Using Bayesian IRT Model}}
    \includegraphics[scale=1]{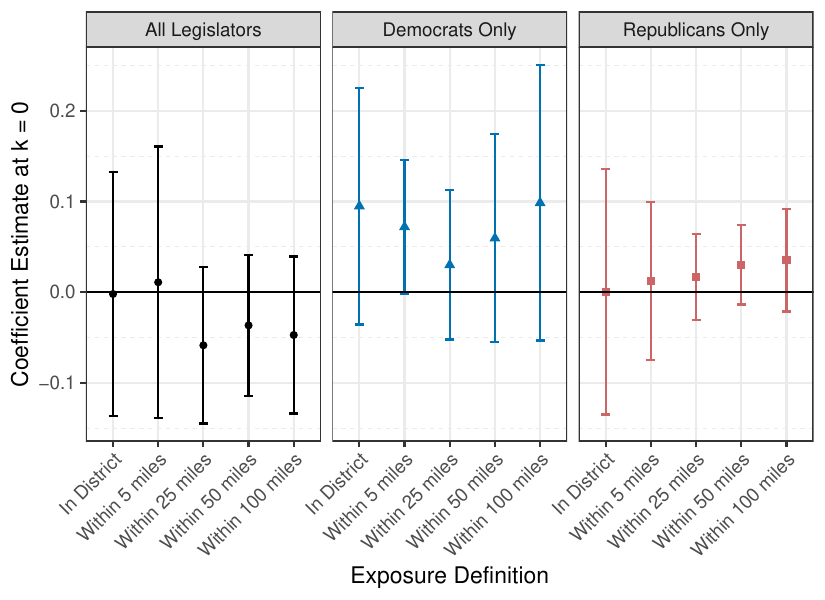}
    \begin{flushleft}
    \scriptsize{\textit{Note}: This figure plots the ATT estimates for the first post-treatment period (\(k = 0\)) pooled across all shootings, when the outcome variable is the dynamic IRT scores. Each panel represents a model subset: all treated legislators v. all untreated legislators, treated Democrats v. untreated Democrats, and treated Republicans v. untreated Republicans. For each panel, we also report \(k = 0\) estimates for the broader treatment assignment definitions. 95\% confidence intervals are based on robust standard errors clustered at the state-, legislator-, chamber-, and year levels.}    
    \end{flushleft}
    \label{fig:irt_did}
\end{figure}

\clearpage

\section{Alternative Measurement Strategies}
\label{app:alternative_results}
\renewcommand{\thefigure}{A.8.\arabic{figure}}
\setcounter{figure}{0}
\renewcommand{\thetable}{A.8.\arabic{table}}
\setcounter{table}{0}

\subsection{Rankings}
\label{app:rankresults}

Following the same DiD design, we report the ATT estimates using the relative rankings of legislators within a state-year --- grouping legislators by state and year and ranking their Gun Issue Scores relative to one another --- in Figure \ref{fig:rank_did}. A positive coefficient indicates that treated legislators move down in ranking (i.e., become relatively less pro-gun control compared to their peers). Similarly, we find null effects after a shooting ($t=0$ and beyond) across all comparisons.

\begin{figure}[h!]
    \centering
    \caption{\textbf{ATT Estimates for Effect of Mass Shooting When Using Ranking}}
    \includegraphics[scale=.9]{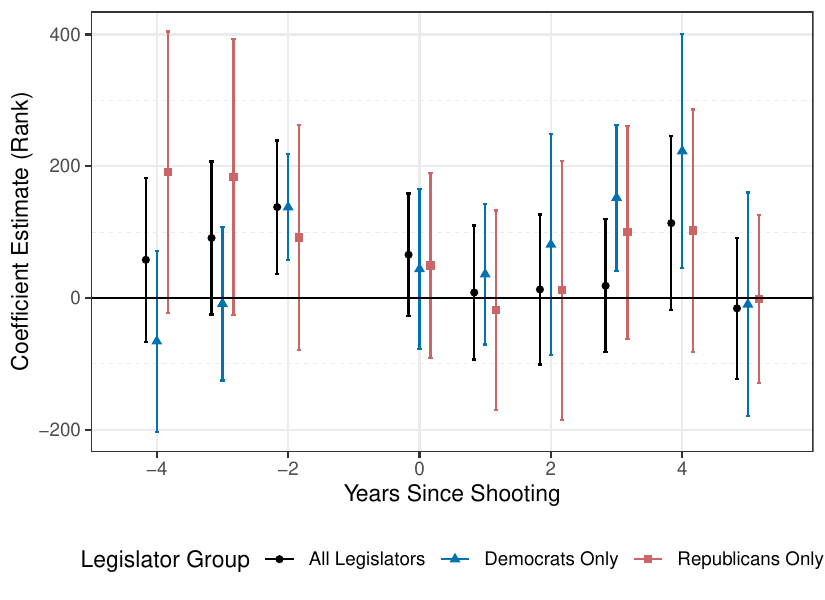}
    \begin{flushleft}
    \scriptsize{\textit{Note}: This figure plots the ATT estimates pooled across all shootings, when the outcome is the within-state-year ranking. Each model represents a different subset: all treated legislators versus all untreated legislators, treated Democrats versus untreated Democrats, and treated Republicans versus untreated Republicans. 95\% confidence intervals are based on robust standard errors clustered at the state-shooting, legislator, chamber, and time period levels.}    
    \end{flushleft}
    \label{fig:rank_did}
\end{figure}

\clearpage

\subsection{NRA Ratings}
\label{app:NRAscores}
As for survey responses from state legislators on firearms policy, the most commonly used source is the National Rifle Association (NRA) scorecards, which rate state legislators' favorability towards firearms policies from self-reported surveys.\footnote{The NRA sent surveys to candidates to assess their alignment with NRA core values, and grade them from a F to A+ letter grade scale.} Upon close examination of the NRA scores, they suffer from severe missingness and inconsistencies in measurement that pose issues that ultimately prevent us from analyzing them.\footnote{See, for example, \url{https://www.thetrace.org/2016/11/nra-gun-record-rating-system-straight-a-students}.} However, as Figure \ref{fig:nra_comparison} shows, the Gun Issue Score tracks more variation in legislators' scoring, as most legislators' NRA ratings stayed unchanged from past elections. At the same time, when a legislator's NRA score changes, we find that the Gun Issue Score also moves in the same direction.

\begin{figure}[!h]
    \centering
    \caption{\textbf{NRA Scoring Compared to Gun Issue Score}}
    \includegraphics[scale=.9]{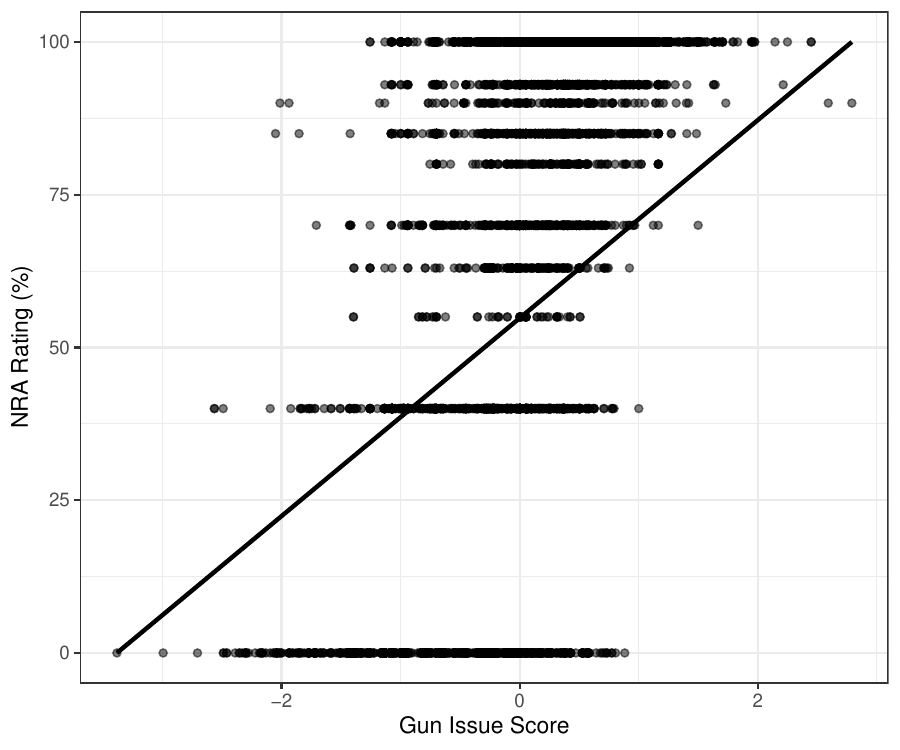}
    \begin{flushleft}
    \scriptsize{\textit{Note}: This figure plots the correlation between the NRA scoring for legislators (where available) and the Gun Issue Score. We convert an A+ rating to a score of 100, an F rating to 0, and scale the intermediate scores proportionally. We successfully matched 5,095 legislators across two datasets. The correlation is 0.58.}
    \end{flushleft}
    \label{fig:nra_comparison}
\end{figure}

\clearpage

\section{Shooting-Specific Estimates}
\label{app:shooting_ests}
\renewcommand{\thefigure}{A.9.\arabic{figure}}
\setcounter{figure}{0}
\renewcommand{\thetable}{A.9.\arabic{table}}
\setcounter{table}{0}

\begin{figure}[h!]
    \centering
    \caption{\textbf{ATT Estimates for Individual Shootings}}
    \includegraphics[scale=0.9]{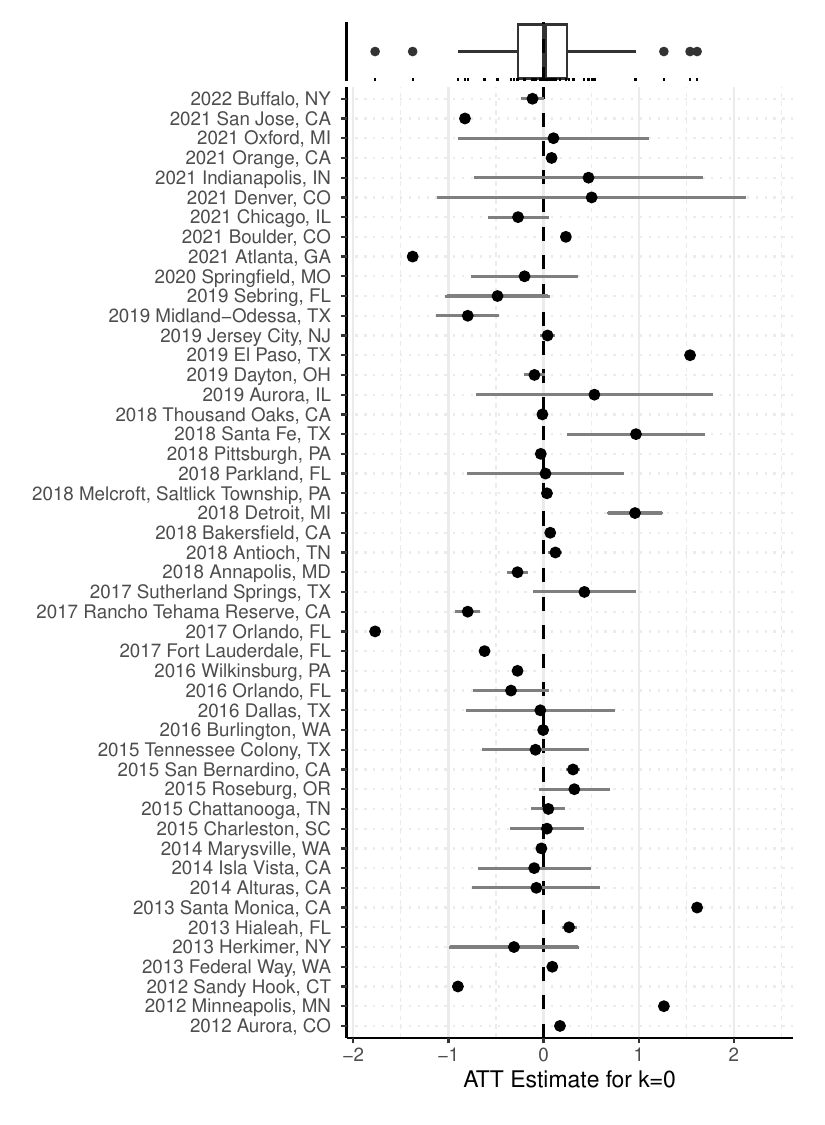}
    \begin{flushleft}
    \scriptsize{\textit{Note}: This figure presents the distribution and individual estimates of the average treatment effect on the treated legislators at $t=0$ for mass shootings included in our overall analyses. The top panel displays the overall distribution of ATT estimates as a boxplot, with individual estimates indicated by rug marks and the dashed vertical line denoting the null (0) effect. The bottom panel shows the corresponding forest plot of ATT estimates for each shooting event, with 95\% confidence intervals where sufficient treated legislator observations exist; estimates based on a single treated legislator are shown without intervals. Positive values indicate an increase in Gun Issue Scores following a mass shooting, while negative values indicate a decrease.}    
    \end{flushleft}
    \label{fig:disagg_ates}
\end{figure}

\clearpage

\section{Sensitivity Analysis for Main Results}
\label{app:sensitivity}
\renewcommand{\thefigure}{A.10.\arabic{figure}}
\setcounter{figure}{0}
\renewcommand{\thetable}{A.10.\arabic{table}}
\setcounter{table}{0}

\subsection{Permutation Tests for Shooting-Specific Estimates}
\label{app:permutation}
Our identification strategy assumes that the Gun Issue Score captures meaningful changes in legislators' voting behavior following shootings. To validate this assumption and assess whether observed effects might arise from random variation, we conduct permutation-based placebo tests. These tests examine whether treatment effect estimates for actual exposure to shootings differ systematically from what would occur if treatment status were randomly assigned.

We perform 100 permutations for each shooting event, creating a null distribution by randomly assigning the same number of ``treated'' legislators as in the actual analysis while preserving the data structure. This approach generates a counterfactual: what effect estimates would we obtain if the timing and location of treatment were arbitrary rather than determined by actual shooting exposure?

The results reveal substantial heterogeneity across shootings, with actual effect estimates ranging from negative to positive values. However, only 6 estimates (12.5\%) fall outside the 95\% range of their corresponding placebo distributions, a rate only marginally above the 5\% false positive rate expected under the null hypothesis of no effect. This pattern suggests that these effects are neither large nor consistent enough to be reliably distinguished from random variation using our permutation test.

This finding does not necessarily imply an absence of true effects but rather indicates that effects are heterogeneous and often modest relative to natural session-to-session variation in legislators' voting patterns. Further, our statistical power to detect shooting-specific effects is limited given the available sample size (1--3 treated legislators per shooting). The permutation tests thus provide important context for interpreting our main findings: individual shooting events rarely produce effects large enough to be statistically distinguishable from chance variation alone.

\begin{figure}[!h]
    \centering
    \caption{\textbf{Comparison of Actual Estimates Against Permutation Distributions}}
    \label{fig:placebo_violin}
    \includegraphics[scale=0.48]{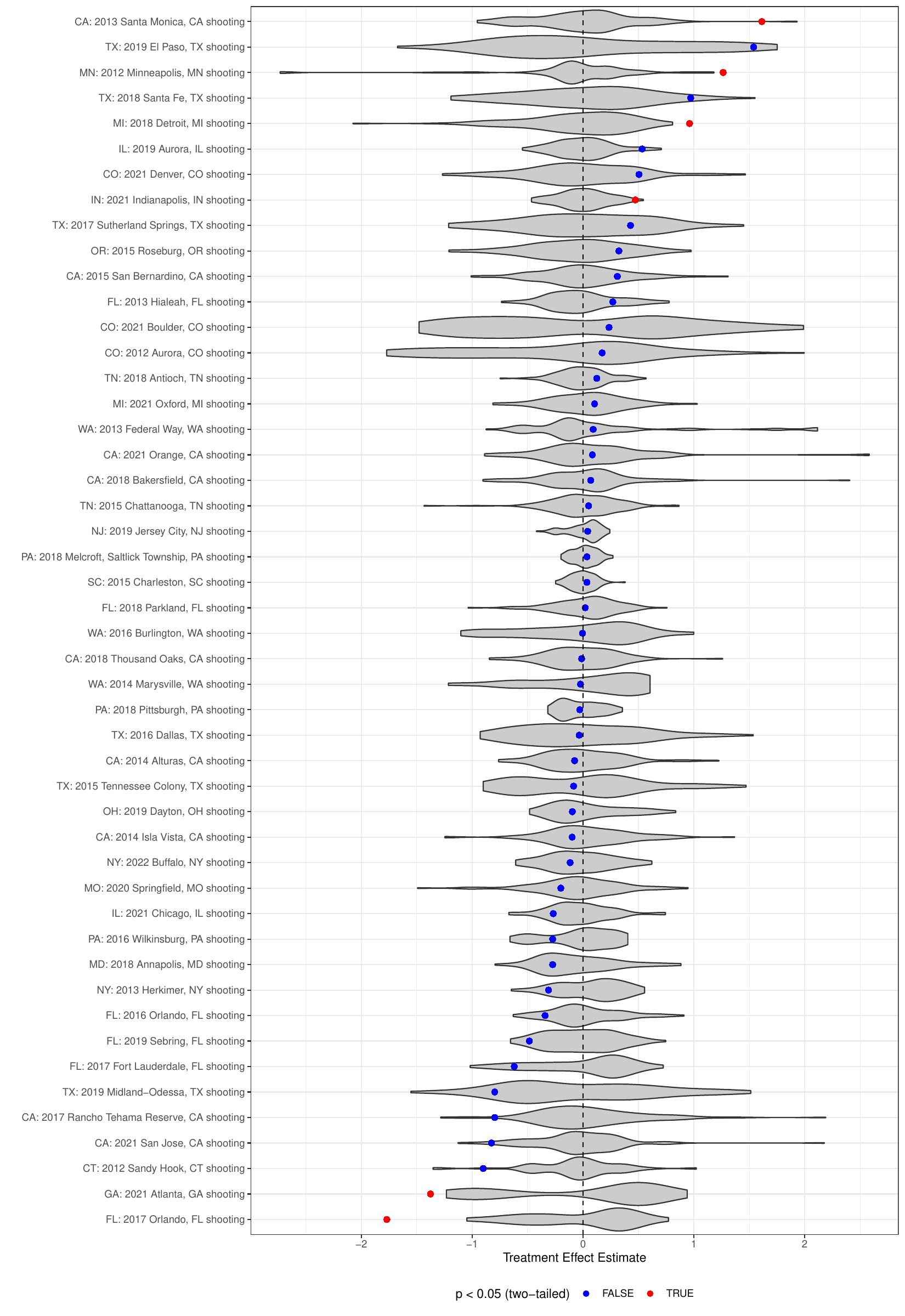}
    \begin{flushleft}
    \scriptsize{\textit{Note}: For each shooting event (rows), violin plots show the distribution of 100 placebo estimates obtained by randomly assigning treatment status to the same number of legislators as actually affected. Points represent actual treatment effect estimates. Red points indicate estimates that fall in the extreme 5\% of their corresponding placebo distribution (two-tailed test). Only 6 of 48 shooting-specific estimates (12.5\%) are statistically significant at the 5\% level, a rate only slightly above what would be expected by chance alone (5\%). The distribution of actual estimates largely overlaps with the placebo distributions, suggesting limited evidence for systematic effects.}    
    \end{flushleft}
\end{figure}

\clearpage

\subsection{Parallel Trends Assumption Holds Overall}
\label{app:parallel_trends_pre_trends}

\begin{figure}[!h]
    \centering
    \caption{\textbf{The Average Effect by Length Of Exposure}}
    \includegraphics[scale=1]{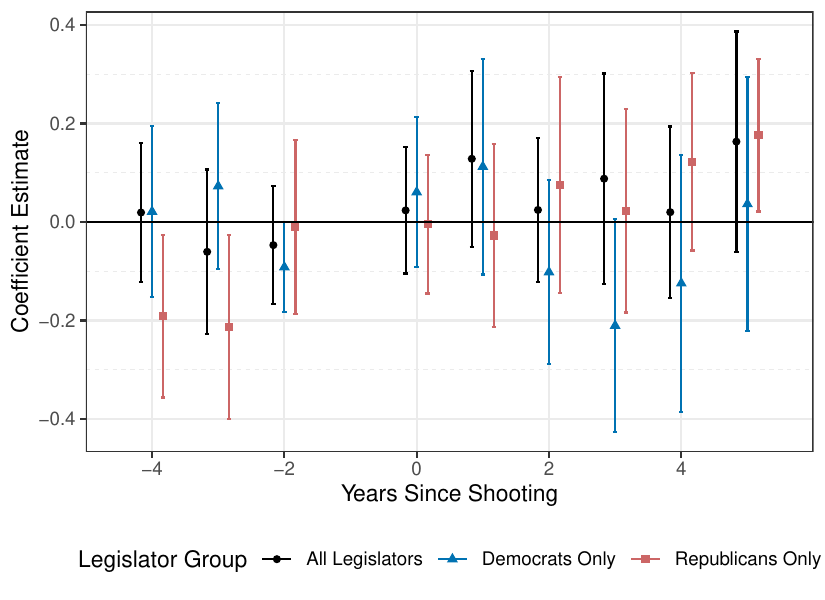}
    \begin{flushleft}
    \scriptsize{\textit{Note}: This figure plots the aggregated ATT at given time periods as a ``dynamic'' specification.}
    \end{flushleft}
    \label{fig:parallel_trends}
\end{figure}

\clearpage

\begin{figure}[!h]
    \centering
    \caption{\textbf{Sensitivity Analysis of the Parallel Trends Assumption Using Smoothness Restrictions}}
    \includegraphics[scale=0.9]{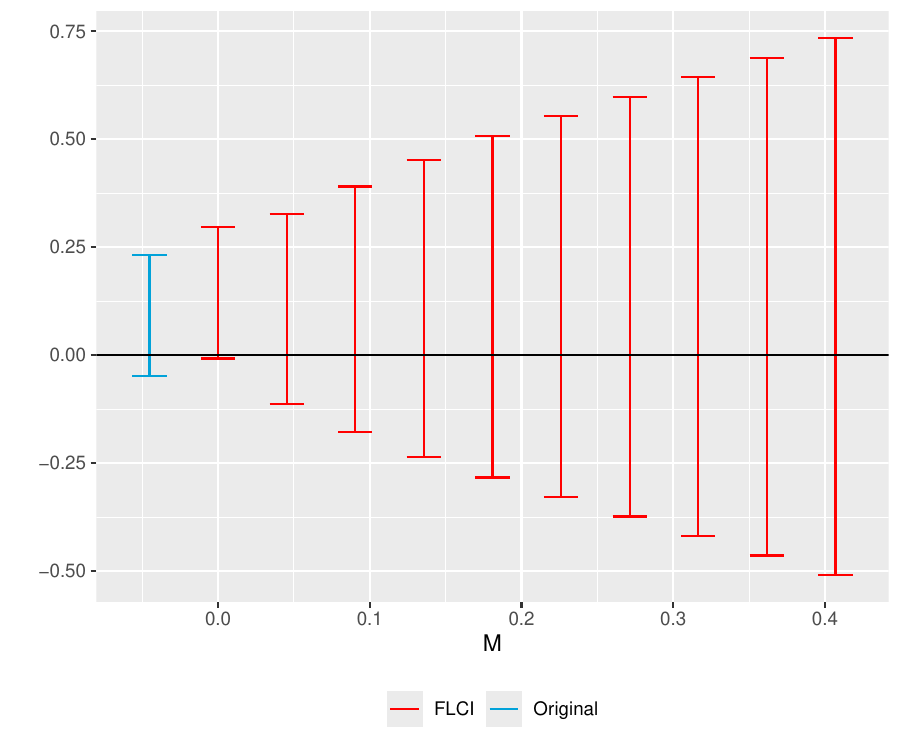}
    \begin{flushleft}
    \scriptsize{\textit{Note}: We implemented the sensitivity analysis using the \texttt{HonestDiD} package under smoothness restrictions \citeApp{rambachan2023more}. We plot the 95\% confidence interval of the in-district treatment effect in blue, and the robust confidence intervals of the treatment effect allowing for non-linearity of the pre-trend up to M in red.}
    \end{flushleft}
    \label{fig:cs_HDiD_smooth}
\end{figure}

Figure \ref{fig:cs_HDiD_smooth} plots the original DiD estimate for the effect of a mass shooting on the Gun Issue Score of state legislators in blue. The red fixed length confidence intervals (FLCIs) report different effect estimates along values of \textit{M}, where \(M = 0\) allows only for linear violations of the parallel trends and larger deviations from linearity are allowed under larger values of \textit{M} \citeApp{rambachan2023more}. The breakdown value is 0: we cannot rule out null effects for the effect of a mass shooting on the Gun Issue Score of state legislators.

\clearpage

\begin{figure}[!h]
    \centering
    \caption{\textbf{Sensitivity Analysis of the Parallel Trends Assumption Using Relative Magnitudes Restrictions}}
    \includegraphics[scale=0.9]{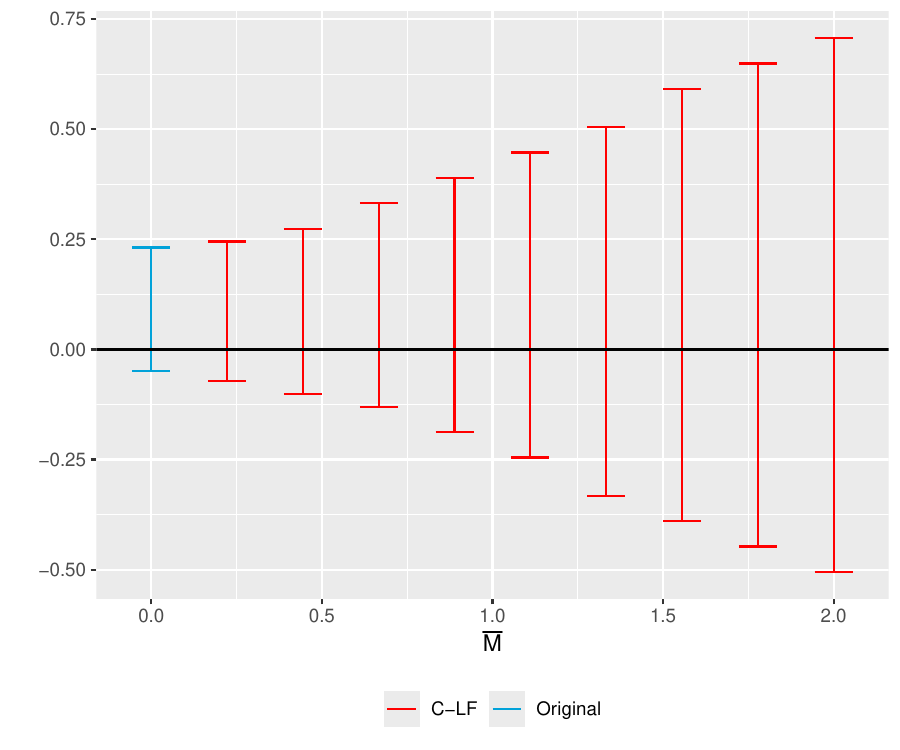}
    \begin{flushleft}
    \scriptsize{\textit{Note}: We implemented the sensitivity analysis using the \texttt{HonestDiD} package under relative magnitudes restrictions \citeApp{rambachan2023more}. We plot the 95\% confidence interval of the in-district treatment effect in blue, and the robust confidence intervals that allows for parallel trends violations up to $\bar{M}$ times the worst pre-treat violation of the parallel trends in red.}
    \end{flushleft}
    \label{fig:cs_HDiD_relmag}
\end{figure}

In Figure \ref{fig:cs_HDiD_relmag}, we see that there is no ``breakdown value’’ \citeApp{rambachan2023more} for a significant effect. Were we to see a significant effect at, say, \(\bar{M} = 1\), this would suggest that a significant result is robust to allowing for violations of parallel trends no larger than the maximum pre-treatment violation. Given the null finding across all values, this suggests that we cannot rule out null results even when allowing for large or small violations of parallel trends.

\clearpage

\section{Assigning Treatment to a District}
\label{app:treatment_district}
\renewcommand{\thefigure}{A.11.\arabic{figure}}
\setcounter{figure}{0}
\renewcommand{\thetable}{A.11.\arabic{table}}
\setcounter{table}{0}

To estimate the effects of a mass shooting on the Gun Issue Score of a district, we first confine our period of study to 2012-2022 to remove any concerns that we are not appropriately accounting for redistricting. Our estimation strategy is similar to those underlying our analyses of individual legislators, though we include chamber- and year-fixed effects and cluster standard errors by chamber and year. The full results can be seen in Figure \ref{fig:app_treat_dist}. 

\begin{figure}[!h]
    \centering
    \caption{\textbf{ATT Estimates for Effect of Mass Shooting on District}}
    \includegraphics[scale=1]{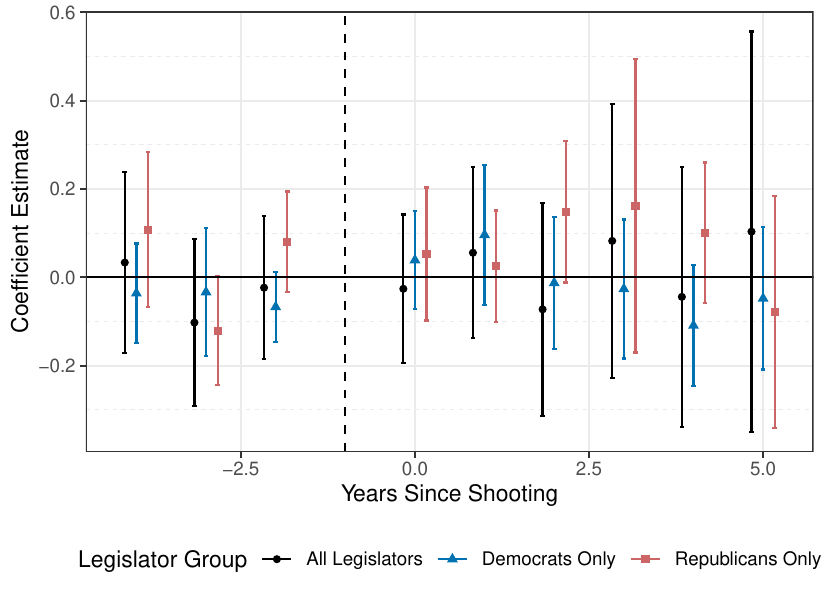}
    \begin{flushleft}
    \scriptsize{\textit{Note}: This figure plots the ATT estimates pooled across all shootings, when treatment is assigned to a district rather than a legislator representing a district. Each panel represents a model subset: all treated legislators v. all untreated legislators, treated Democrats v. untreated Democrats, and treated Republicans v. untreated Republicans. For each panel, we also report \(k = 0\) estimates for the broader treatment assignment definitions. 95\% confidence intervals are based on robust standard errors clustered at the state-, chamber-, and year levels.}
    \end{flushleft}
    \label{fig:app_treat_dist}
\end{figure}

\clearpage

%---------------------------
\begingroup
\setstretch{1}
\bibliographystyleApp{apsr}
\bibliographyApp{bib}
\endgroup

\endgroup

\end{document}